%% file: wileyNJD-Doc.tex
\PassOptionsToPackage{symbol}{footmisc}
\documentclass[AMA,STIX1COL]{WileyNJD-v2}
\usepackage{moreverb}
\usepackage[frozencache=true,cachedir=minted-cached-files]{minted}
\usepackage[short]{optidef}
\usepackage{subcaption}
\usepackage{pgfplots}

\pgfplotsset{compat=newest}
\usetikzlibrary{decorations,snakes,positioning,
  calc,
  intersections,
  fit}
\usetikzlibrary{shapes.arrows}
\usepgfplotslibrary{patchplots}
\pgfdeclarelayer{background}
\pgfdeclarelayer{layer1}
\pgfsetlayers{background,layer1,main}
\usepackage{relsize}  % needed by \mathsmaller, to make nice ^T (transpose)
\newcommand{\inputpdf}[1]{%
  \includegraphics{pdf_figures/#1.pdf}%
}

\newcommand\BibTeX{{\rmfamily B\kern-.05em \textsc{i\kern-.025em b}\kern-.08em
T\kern-.1667em\lower.7ex\hbox{E}\kern-.125emX}}

\renewcommand{\thefootnote}{\fnsymbol{footnote}}

\newcommand{\inRpow}[1]{\in \mathbb{R}^{#1}}

\newcommand{\vb}[1]{\ensuremath{\textbf{#1}}}
\newcommand{\Tr}{\ensuremath{\mathsf{\mathsmaller T}}} 
\newcommand{\muaompc}{\ensuremath{\mu}\emph{AO-MPC}}

\newlength\figHsmall
\newlength\figWsmall
\newlength\figHmedium
\newlength\figWmedium
\newlength\figHbig
\newlength\figWbig
\input{colors.tex}
\articletype{Article Type}%

\received{<day> <Month>, <year>}
\revised{<day> <Month>, <year>}
\accepted{<day> <Month>, <year>}

\begin{document}

\title{Flexible development and evaluation of machine-learning-supported optimal control and estimation methods via HILO-MPC}

\author[1]{Johannes Pohlodek}%$^{\dagger,}$}

\author[1]{Bruno Morabito$^{\ddagger,}$}%$^{\dagger,\ddagger,}$}

\author[1]{Christian Schlauch}

\author[2]{Pablo Zometa}

\author[3]{Rolf Findeisen$^{\ast,\ddagger,}$}

\authormark{Pohlodek \textsc{et al}}

\address[1]{\orgdiv{Laboratory for Systems Theory and Automatic Control}, \orgname{Otto~von~Guericke~University~Magdeburg}, \orgaddress{\country{Germany}}}

%\address[2]{\orgdiv{International Max Planck Research School (IMPRS) for Advanced Methods in Process and Systems Engineering}, \orgaddress{\state{Magdeburg}, \country{Germany}}}

\address[2]{\orgname{German~International~University~Berlin}, \orgaddress{\country{Germany}}}

\address[3]{\orgdiv{Control and Cyber-Physical Systems Laboratory}, \orgname{TU~Darmstadt}, \orgaddress{\country{Germany}}}

\corres{*Rolf Findeisen, Control and Cyber-Physical Systems Laboratory, TU~Darmstadt, Germany. \email{rolf.findeisen@tu-darmstadt.de}}

% \presentaddress{Present address}

\abstract[Abstract]{Model-based optimization approaches for monitoring and control, such as model predictive control and optimal state and parameter estimation, have been used successfully for decades in many engineering applications. Models describing the dynamics, constraints, and desired performance criteria are fundamental to model-based approaches. 
Thanks to recent technological advancements in digitalization, machine learning methods such as deep learning, and computing power, there has been an increasing interest in using machine learning methods alongside model-based approaches for control and estimation. The number of new methods and theoretical findings using machine learning for model-based control and optimization is increasing rapidly. However, there are no easy-to-use, flexible, and freely available open-source tools that support the development and straightforward solution to these problems. This paper outlines the basic ideas and principles behind an easy-to-use Python toolbox that allows to quickly and efficiently solve machine-learning-supported optimization, model predictive control, and estimation problems. The toolbox leverages state-of-the-art machine learning libraries to train components used to define the problem. It allows to efficiently solve the resulting optimization problems. Machine learning can be used for a broad spectrum of problems, ranging from model predictive control for stabilization, setpoint tracking, path following, and trajectory tracking to moving horizon estimation and Kalman filtering. For linear systems it enables quick generation of code for embedded MPC applications. HILO-MPC is flexible and adaptable, making it especially suitable for research and fundamental development tasks. Due to its simplicity and numerous already implemented examples, it is also a powerful teaching tool. The usability is underlined, presenting a series of application examples.}
%, spanning from the problem to learn the dynamic model of a car for racing, learning references to be tracked, learning the solution of an MPC controller itself, and the embedded implementation learning-based MPC controller.}


%\jnlcitation{\cname{%
%\author{J. Pohlodek}, 
%\author{B. Morabito}, 
%\author{C. Schlauch},
%\author{P. Zometa}, and 
%\author{R. Findeisen}}, 
%\ctitle{Flexible development and evaluation of machine learning supported optimal control and estimation methods via HILO-MPC}, \cjournal{Int J Robust Nonlinear Control}, \cvol{2017;00:1--6}.}

\maketitle

%\footnotetext[1]{Both authors contributed equally.}
\footnotetext[2]{BM and RF are affiliated with the International Max Planck Research School (IMPRS) for Advanced Methods in Process and Systems Engineering, Magdeburg.} 
\footnotetext[0]{The authors acknowledge funding of the DIGIPOL project (Magdeburg, Saxony-Anhalt) funded in the EU-ERDF scheme, and in the frame of the LOEWE initiative (Hesse, Germany) within the emergenCITY center.}
\renewcommand*{\thefootnote}{\arabic{footnote}}

\section{Introduction}
Advanced optimal control and estimation problems such as model predictive control, moving horizon estimation, and Kalman filters require models representing, e.g., the system dynamics, constraints, reference values, and objective functions. Based on these models, the inputs,  the estimates of the states, or parameters are found based on the minimization of an objective function \cite{brunton2019data,hou2013model}
Often obtaining accurate models is challenging. For example, dynamical models can be constructed with first principles, such as conservation of energy, matter, and thermodynamics laws. 
However, underlying unknown nonlinearities, parameters, or coupling effects, often results in complex dynamics, which is challenging to model, thus rendering it difficult to employ such model-based methods for process control or numerical simulations. 

Alternatively, if only data is used to derive a model, one generally talks about data-driven or machine learning models.\cite{brunton2019data,glassey2018hybrid}  Machine learning has recently obtained much attention since it allows to model and to simulate complex, not well-understood phenomena or efficiently simulate complex, large-scale systems, provided that a sufficiently large amount of data is available. For many decades the scarcity of data, features selection, and computational power have been the main bottlenecks to the widespread use of machine learning for modeling and control. The situation is rapidly changing due to the rapid digitization and interconnection (e.g., Internet of Things and Industry 4.0 initiatives), advances in machine learning, and cloud computing.\cite{baheti2011cyber,lucia2016predictive} Hence, machine learning is finding more and more applications, not only in research but in everyday life. In recent years, the number of machine learning applications, publications, new methods, and theoretical findings has increased rapidly. Furthermore, open-source libraries for machine learning such as TensorFlow\cite{tensorflow2015} and PyTorch\cite{pytorch} offer a low entry barrier to this technology and allow fast development of machine learning solutions. The combination of these factors boosted the quality and quantity of results in many fields such as medicine,\cite{Shen2017} autonomous driving,\cite{Arnold2019} transportation,\cite{Nguyen2018} Internet of Things,\cite{Mohammadi2018} image recognition\cite{Voulodimos2018} and many others. The control and optimization community also recognized the usefulness of machine learning for control, e.g., for efficient and safe optimal control and model predictive control (MPC), motivating many contributions.\cite{hou2013model,Hewing2020,maiworm2021online,aswani2013provably}
%, see \cite{hou2013model,Hewing2020,maiworm2021online,aswani2013provably} and references therein.

\begin{figure*}[h]
    \centering
    \inputpdf{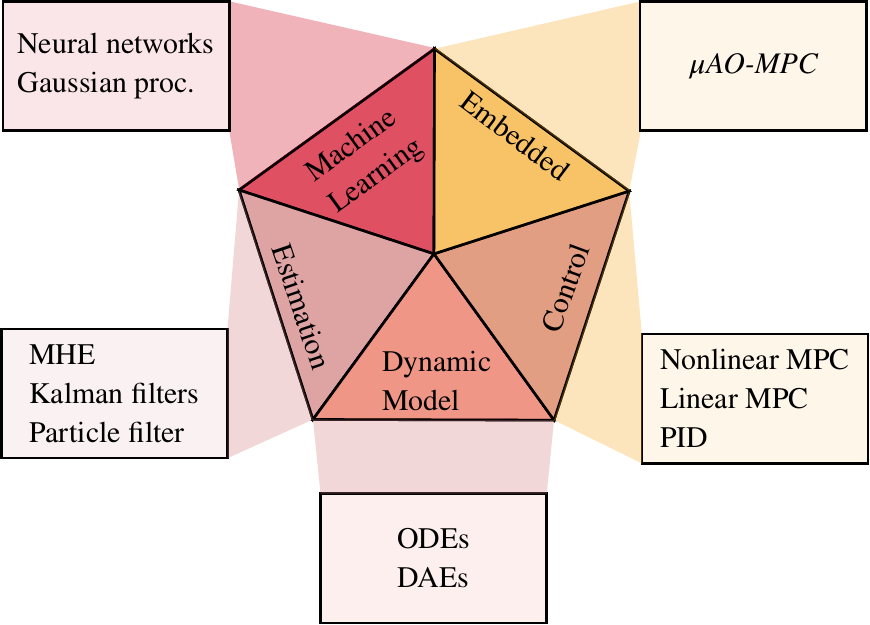}
    \caption{Core elements of HILO-MPC.}
    \label{fig:overview}
\end{figure*}

\begin{figure*}[htb]
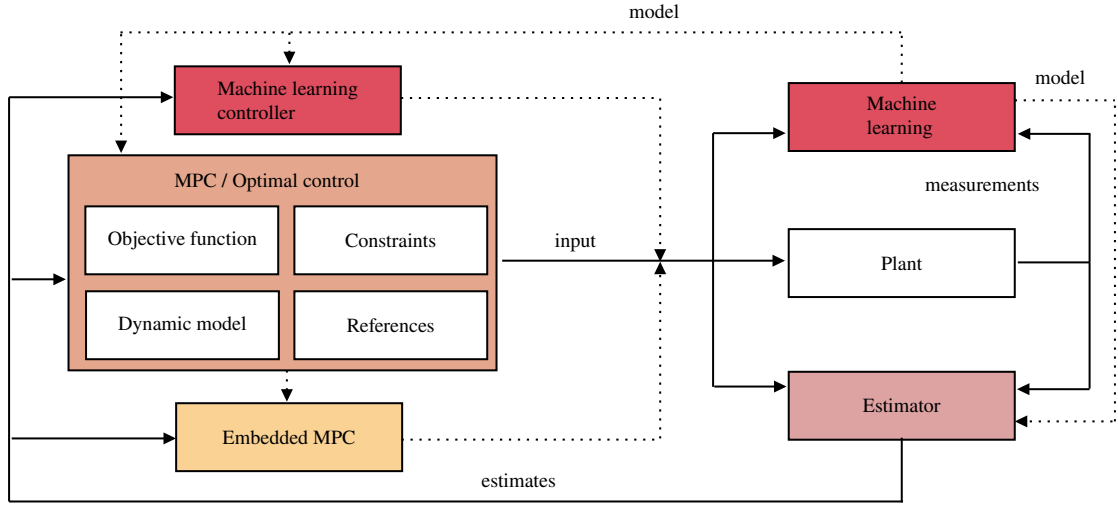

    \centering
\include{figures/control_loop}
\caption{Schematic diagram of the interactions between the components of the toolbox. The machine learning component can be used to learn model parts (see Section~\ref{sec:example-race-car}), cost functions, constraints, or references/disturbances (see Section~\ref{sec:cooperative-robots}) from real or simulation data, that can then be used in the controller or estimator. It can also be used to learn an approximated solution of an optimal controller/MPC controller (see Section~\ref{sec:spring-damper}). It is also possible to generate embedded code for the controller (see Section~\ref{sec:embedded}). In principle, similarly, the same holds for learning components of the estimator.}
    \label{fig:scheme}
\end{figure*}

While machine learning libraries are powerful tools to build machine learning models, using these models in a control strategy requires typically ad-hoc solutions, which are complicated, prone to bugs, and not easily transferable to other applications. In the frame of this work, we outline an approach to interconnect machine learning methods and tools with methods from optimization-based control and estimation using Python. The resulting open-source Python library, named HILO-MPC \footnote{HILO-MPC stands for mac{\bf HI}ne {\bf L}earning and {\bf O}ptimization for {\bf M}odeling, {\bf P}rediction and {\bf C}ontrol.} aims at providing a way to easily implement machine learning models into a wide variety of control, optimization and estimation problems, tailored towards fundamental research and basic development tasks (see Fig.~\ref{fig:overview}). It interfaces with PyTorch and TensorFlow to train neural networks and uses in-house code to train Gaussian processes. The trained models can be used in a large variety of control or estimation problems for example, in the system's dynamics, the constraints, the cost functions, or even the controller itself (see Figure~\ref{fig:scheme}).

To speed up the development time, HILO-MPC provides a way to define and solve a broad spectrum of optimal control and estimation problems quickly and with minimum effort. Table~\ref{tab:problems-hilo} summarizes the different problems that can be solved. The philosophy followed is the ease of use and modularity. Ease of use is achieved by providing an intuitive high-level interface for the problem definition. At the same time, modularity is obtained by ensuring effortless interfaces between the different tools (see Figure~\ref{fig:scheme}). This allows using the toolbox both for research and for teaching. Table \ref{tab:problems-hilo} summarizes the main tools and methods available.

The main contribution of this paper is the presentation of a modular way to fuse modeling, simulation, estimation, control, machine learning approaches, and optimization, resulting in a flexible yet easy-to-use open-source toolbox. The usability is underlined, presenting a series of application examples, spanning from the problem to learn the dynamic model of a car for racing, learning references to be tracked, learning the solution of an MPC controller, and the embedded implementation of a MPC controller. Note that, while we focus on the integration of machine learning into control and estimation problems, most of the parts can also be used without any machine learning component. 

The remainder of this paper is structured as follows. Section~\ref{sec:hilo-mpc} provides a brief overview of toolboxes available for optimal and predictive control, with a specific focus on open-source toolboxes.

In Section~\ref{sec:hilo-mpc} the different modules of the toolbox are described. Furthermore, we show example code snippets to illustrate the required syntax. In Section~\ref{sec:examples} we apply the toolbox in four examples. Finally, in Section~\ref{sec:conclusions} we conclude and give an outlook on current and future developments.% of the toolbox.

Note that this paper only provides a glimpse of the principles and main ideas and the usability via some example applications. It is not a manual, as providing a comprehensive description of the toolbox is outside the scope of this work. For this, the reader is referred to the documentation.\footnote{The full documentation and toolbox are available under \url{http://www.ccps.tu-darmstadt.de/hilo_mpc}.}
\begin{table*}[b]
    \caption{Components and formulations imtegrated in HILO-MPC.}
    \label{tab:problems-hilo}
    \centering
    \begin{tabular}{lllll}
        \toprule
        Models & Controllers & Machine Learning & Estimators & Embedded \\
        \midrule
        \parbox[c][2.5cm][c]{3.5cm}{Linear/nonlinear\\Time-invariant/variant\\Continuous/discrete\\ODEs, DAEs} & \parbox[c][2.5cm][c]{3.5cm}{Nonlinear MPC\\Trajectory tracking MPC\\Path-following MPC\\PID} & \parbox[c][2.5cm][c]{3cm}{Neural networks\\Gaussian processes} & \parbox[c][2.5cm][c]{3.9cm}{Moving horizon estimation\\Kalman filter\\Extended Kalman filter\\Unscented Kalman filter\\Particle filter} & \muaompc\cite{zometa2013} \\
        \bottomrule
    \end{tabular}
\end{table*}

%%%%%
\section{A Glimpse Review of Open-source MPC tools}
\label{sec:OSMPC}
% We do not aim to provide a complete review of toolboxes available for model predictive and optimal control. We rather focus on open-source available tools comparing their connectivity and interrelation to machine learning tools.

\begin{table*}[htb]
    \caption{Comparison of different freely accessible MPC tools. ML: machine learning, MHE: moving horizon estimator, LMPC: linear MPC, PFMPC: path-following MPC, KF: Kalman filters, PF: particle filters}
    \label{tab:comparison-toolboxes}
    \centering
    \begin{tabular}{llclll}
        \toprule
        & Problems & ML & Interface & Focus & Article \\
        \midrule
        ACADO & MPC & No & C++, MATLAB & Embedded & Houska et al.\cite{Houska2011a} \\
        ACADOS & MPC, MHE & No & \parbox[c][1cm][c]{3.2cm}{C, MATLAB, Octave,\\Python, Simulink} & Embedded & \parbox[c][1cm][c]{2.8cm}{Verschueren et al.\cite{Verschueren2018}\\Verschueren et al.\cite{Verschueren2019}} \\
        do-mpc & \parbox[c][1cm][c]{3.2cm}{MPC, MHE,\\Multi-stage MPC} & No & Python & Multi-stage MPC & Lucia et al.\cite{LUCIA201751} \\
HILO-MPC & \parbox[c][1cm][c]{3.2cm}{MPC, PFMPC, LMPC\\MHE, KF, PF} & Yes & Python & Machine Learning & \\
        MPCTools & MPC, MHE & No & \parbox[c][1cm][c]{3.2cm}{Python,\\Octave (MATLAB)} & MPC, MHE & \parbox[c][1cm][c]{3.4cm}{Risbeck and Rawlings\cite{risbeck2015nonlinear}\\Risbeck and Rawlings\cite{risbeck2016nonlinear}} \\
        MPT3 Toolbox & LMPC & No & MATLAB & Comp. Geometry & Herceg et al.\cite{MPT3} \\
        YALMIP & MPC, MHE & No & MATLAB & Optimization & L\"{o}fberg\cite{YALMIP} \\
        \bottomrule
    \end{tabular}
\end{table*}

%Comparing HILO-MPC with other toolboxes is challenging since it can solve a wide variety of control and estimation problems. 
By now a wide variety of toolboxes for MPC exist. We ocus on freely available, open-source toolboxes. The software ACADO\cite{Houska2011a} and ACADOS\cite{Verschueren2018} aim at efficiency and fast solution keeping embedded applications in mind. Note that, although interfaces to embedded MPC libraries are offered, embedded MPC is not the main focus of the toolbox. MPT3 focuses on computational geometry, which is particularly useful for some classes of robust MPC approaches.\cite{MPT3} YALMIP instead offers a broad spectrum of optimization problems, such as bilevel optimization and mixed-integer programming, and some MPC schemes.\cite{YALMIP} The toolbox do-mpc implements robust multi-stage MPC and moving horizon estimation (MHE).\cite{LUCIA201751} MPCTools offers some interfaces that help build MPC and MHE control problems in CasADi.\cite{risbeck2015nonlinear,risbeck2016nonlinear} Table~\ref{tab:comparison-toolboxes} summarizes some of the main differences between existing open-source toolboxes. 

Using machine learning models in the previously mentioned toolboxes is often not straightforward. We tackle this challenge by providing wrappers for machine learning libraries and some in-house learning libraries. Furthermore, compared to the other toolboxes, it focuses on a simple and intuitive interface that makes it easy to use also for beginners.

\section{Basic Concepts and Components}
\label{sec:hilo-mpc}
% The core element of HILO-MPC are models, which allow to describe the system dynamics, constraints, cost functions, etc.

To provide flexibility five main components are tightly integrated --- in the following denoted as \emph{modules} named \emph{dynamic model}, \emph{control}, \emph{estimator}, \emph{machine learning}, and \emph{embedded} (see Figure~\ref{fig:overview}). The dynamic model module is used to generate dynamic models which are used in all other modules. The control module allows to setup and formulate controllers, solve and apply them. It focuses, but is not limited to predictive and optimal controllers. The estimator module focuses on  state and parameter estimation using the dynamic models. The machine learning module is responsible for defining and training machine learning models that can be used in any of the previous modules. Finally, the embedded module currently interfaces with embedded linear MPC software (\muaompc\cite{zometa2013}) that allows to easily deploy the code for embedded applications. In the next sections, we will briefly describe the different modules in more detail.\footnote{New tools will be added, and current tools might be modified, please refer always to the current documentation for the newest features..}

The backbone of HILO-MPC is CasADi.\cite{Andersson2019} CasADi is a flexible and widely used tool for algorithmic differentiation and optimization. It has interfaces with a wide range of solvers for linear and quadratic programming (CLP,\cite{john_forrest_2022_5839302} qpOASES\cite{Ferreau2014}), nonlinear programming such as IPOPT,\cite{wachter2006implementation} quadratic mixed-integer programming (CPLEX,\cite{cplex2009v12} GUROBI\cite{gurobi}), for non quadratic nonlinear mixed-integer problems (BONMIN,\cite{bonami2007bonmin} KNITRO\cite{nocedal2006knitro}) and large nonlinear problems WORHP.\cite{nikolayzik2010nonlinear}

CasADI is used to build the models, optimization, and estimation problems. TensorFlow or PyTorch are interfaced by automatically translating the models into CasADi-compatible structures. Hence, a CasADi problem is defined and solved. 

In the following, we briefly outline the core modules and provide insights in the formulations and solution approaches.

\subsection{Dynamic Model Module}
\label{sec:modeling_module}

The dynamic model module forms the foundation for all other components. It is a high-level interface to generate representations of dynamical systems that can be used for model-based controllers and estimators, like MPC or MHE, or inside simulations to reproduce the behavior of a plant (or a controller) in a control loop. The module currently supports linear and nonlinear, time-invariant and time-variant, discrete-time systems described in discrete time by difference equations, continuous-time systems described by ordinary differential equations (ODEs), and differential algebraic equations (DAEs).

For example, the module allows to describe a general time-variant continuous-time nonlinear model given by a DAE of the form
\begin{align}
    \dot{x}(t)&=f(t,x(t),z(t),u(t),p),\nonumber\\
    0&=q(t,x(t),z(t),u(t),p),\label{eq:nonlin_model}\\
    y(t)&=h(t,x(t),z(t),u(t),p)\nonumber.
\end{align}
Here $x(t)\in\mathbb{R}^{n_x}$ is the differential state vector, $z(t)\in\mathbb{R}^{n_z}$ the vector of algebraic states, $u(t)\in\mathbb{R}^{n_u}$ is the input vector and $p\in\mathbb{R}^{n_p}$ is a vector of parameters. The function $f\colon\mathbb{R}\times\mathbb{R}^{n_x}\times\mathbb{R}^{n_z}\times\mathbb{R}^{n_u}\times\mathbb{R}^{n_p}\rightarrow\mathbb{R}^{n_x}$ represents the ODEs of the model, and the function $q\colon\mathbb{R}\times\mathbb{R}^{n_x}\times\mathbb{R}^{n_z}\times\mathbb{R}^{n_u}\times\mathbb{R}^{n_p}\rightarrow\mathbb{R}^{n_z}$ describes the algebraic equations forming a semi-explicit DAE system overall.\footnote{Note that HILO-MPC only supports semi-explicit index 1 DAE systems.\cite{kunkel2006differential} Index 1 indicates DAE systems that can be transformed into a pure ODE system by differentiating the algebraic equations once.} The function $h\colon\mathbb{R}\times\mathbb{R}^{n_x}\times\mathbb{R}^{n_z}\times\mathbb{R}^{n_u}\times\mathbb{R}^{n_p}\rightarrow\mathbb{R}^{n_y}$ describes the measurement equations mapping to some measurable quantity of the system. If these measurement equations are not defined during the setup of the model, the controllers and estimators using the model assume that all states are measured.

When necessary, continuous-time models can be easily discretized using the \mintinline{python}{model.discretize(...)} method. Available discretization methods are explicit Runge-Kutta methods up to the 4th order and implicit collocation schemes.\cite{Biegler2010} For the explicit Runge-Kutta methods, a series of Butcher tableaus is available, while the implicit collocation schemes only support Gauss-Legendre polynomials and Gauss-Radau polynomials for the calculation of the collocation points.

For some applications and control and estimation methods a linear model might be necessary. The direct creation of linear models is also possible by supplying the required matrices during model setup. If only a nonlinear model is available, this can be easily linearized with respect to an equilibrium point using \mintinline{python}{model.linearize(...)}.

\begin{figure*}[htb]
    \captionsetup[figure]{justification=centering,font=large}
    \centering
    \inputpdf{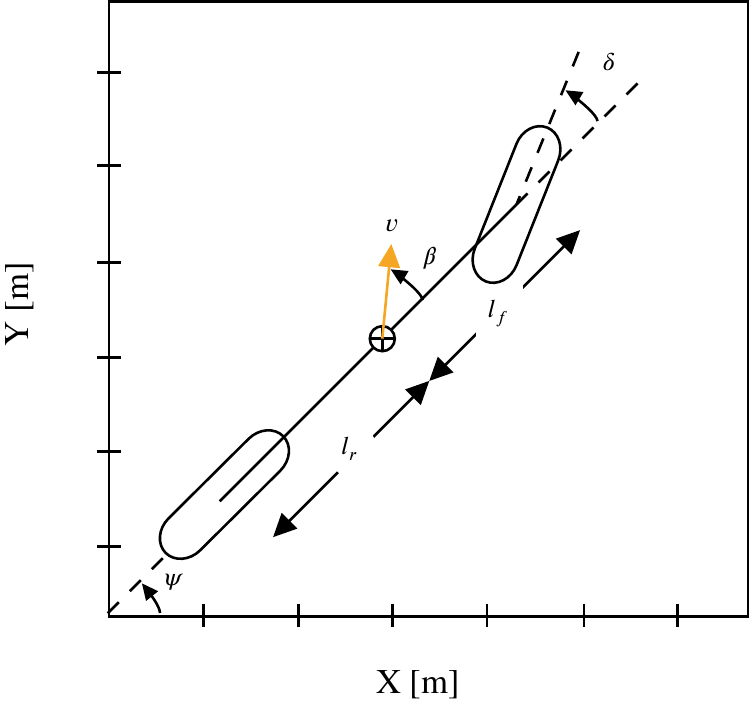}
    \caption{Simple single-track/bicycle car model.}
    \label{fig:bike_simple}
\end{figure*}

To underline the simplicity of defining a model, we consider a simple single-track car/bicycle (see Fig.~\ref{fig:bike_simple}) model which will be used as an example model in the following sections
\begin{align*}
    \dot{p}_\text{x} &= v \cos(\phi(t) + \beta),  \\
    \dot{p}_\text{y} &= v \sin(\phi(t) + \beta), \\
    \dot{v} &= a, \\
    \dot{\phi} &= v/l_\text{r} \sin(\beta), \\
    \beta & = \arctan\left(\frac{l_r}{l_\text{r} + l_\text{f}}\tan(\delta)\right).
\end{align*}
Here, $p_{\text{x}}$ and $p_{\text{y}}$ are the $x$- and $y$-coordinates of the vehicle center of mass, $v$ is the norm of the velocity of the center of mass, and $\phi$ is the orientation angle of the vehicle with respect to the $x$-coordinate. The inputs are the acceleration of the center of mass $a$ and the steering angle $\delta$. The parameter $l_\text{r}$ is the distance between the center of mass and the rear wheel, and $l_\text{f}$ is the distance between the center of mass and front wheel. The corresponding model can be set up as follows
\begin{minted}[bgcolor=bg,breaklines]{python}
from hilo_mpc import Model

# Initialize empty model
model = Model(name='Bike')
# Define model equations
equations = """
# ODEs
dpx/dt = v(t)*cos(phi(t) + beta)
dpy/dt = v(t)*sin(phi(t) + beta)
dv/dt = a(k)
dphi/dt = v(t)/lr*sin(beta)
# Algebraic equations
beta = arctan(lr/(lr + lf)*tan(delta(k)))
"""
model.set_equations(equations=equations)
# Sampling time in seconds 
dt = 0.01 
model.setup(dt=dt)
model.set_initial_conditions(x0=[0, 0, 0, 0])
\end{minted}
Note that the model equations are supplied in the form of a string. HILO-MPC is able to parse the most common functions and operations in this format, while it also provides the flexibility to use other input formats, e.g., by explicitly defining the inputs and states of the system.
% The model can be also defined using the model variables. This reads
% \begin{minted}[bgcolor=bg,breaklines]{python}
% # Initialize empty model
% model = Model(name='Bike')

% # Set the variables
% inputs = model.set_inputs(['a', 'delta'])
% states = model.set_states(['px','py','v','phi']

% # Unwrap states
% px = states[0]
% py = states[1]
% v = states[2]
% phi = states[3]

% # Unwrap states
% a = inputs[0]
% delta = inputs[1]

% # Parameters
% lr = 1.4  # [m]
% lf = 1.8  # [m]
% beta = ca.arctan(lr / (lr + lf) * ca.tan(delta))

% # ODE
% dpx = v * ca.cos(phi + beta)
% dpy = v * ca.sin(phi + beta)
% dv = a
% dphi = v / lr * ca.sin(beta)

% # Pass the differential equations to the model
% model.set_differential_equations([dpx, dpy, dv, dphi])

% # Sampling time in seconds 
% dt = 0.01 
 
% # Setup the model
% model.setup(dt=dt)
 
% # Pass the initial conditions
% model.set_initial_conditions(x0=[0,0,0,0])
% \end{minted}
Units, labels and descriptions of all variables can also be set for convenience and for use in the plotting functionality. For a more in-depth description of available functionalities, refer to the documentation.
%{\color{red}[REFERENCE TO LINK HERE]}
Once the initial conditions are set, the model can be used for simulation by running \mintinline{python}{model.simulate(u=u, p=p)}. Note that, for every integration interval given by the sampling time $\mathrm{d}t$ the input for the continuous times system is kept constant, as it is commonly done.

%% TODO: Move to control part
% Another way to run simulations in combination with controllers and observers in a simple control loop is to use the class \mintinline{python}{ControlLoop} that is specifically designed for this purpose.
One can easily store, access, and plot solutions generated by simulations. Every time the method \mintinline{python}{model.simulate(...)} is executed, the solution is saved in the \mintinline{python}{model.solution} object. This can be accessed similarly to a Python dictionary. To plot the solution, i.e., the evolution of states, inputs, and parameters with time, the user can use the method \mintinline{python}{model.solution.plot()}.

\subsection{Machine Learning Module}\label{sec:learning}

The machine learning module is responsible for defining and training machine learning models. Machine learning models are data-driven models that are used to map selected inputs $v \in \mathbb{R}^{n_v}$, also called features, to outputs $l \in \mathbb{R}^{n_l}$, also called labels. The machine learning models can be used to describe dynamical systems, unknown functions, constraints, cost functions, or a controller (see Fig.~\ref{fig:scheme}).

Two machine learning approaches/models are currently integrated: artificial feedforward fully connected neural networks and Gaussian processes. Further approaches are under development.

\subsubsection{Artificial Neural Networks}
We briefly review artificial neural networks (ANNs) to introduce the required notation and concepts.
\begin{figure}
    \centering
    \inputpdf{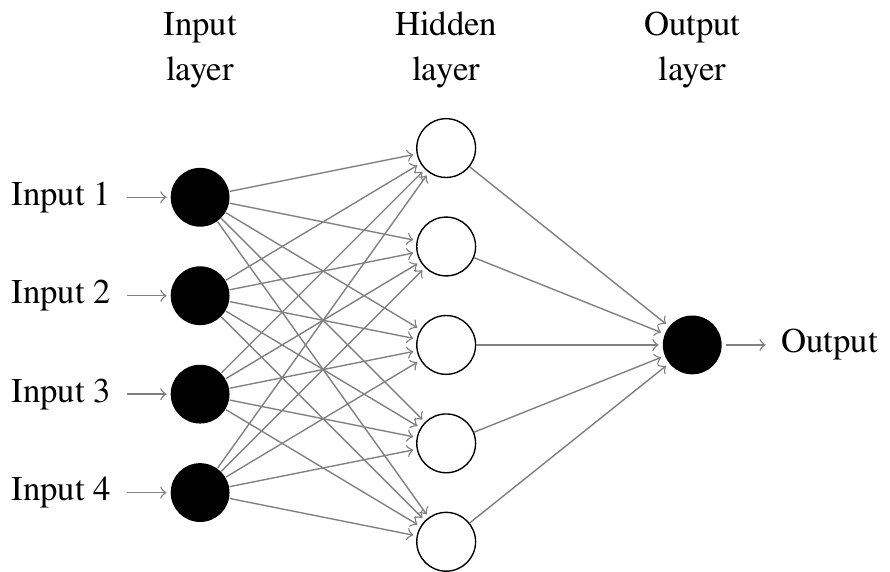}
    \caption{Simple feedforward neural network with four features (inputs), one hidden layer and one label (output).}
    \label{fig:ann}
\end{figure}
Inspired by the neural circuits in brains,\cite{McCulloch1943} ANNs can be represented as a collection of interconnected nodes, the so-called neurons, that are usually organized into layers. We focus on feedforward fully connected ANNs (Figure~\ref{fig:ann}). Each layer of the ANN is mathematically represented by% a linear regression propagated through some function called \emph{activation function}
\begin{equation}
    z_i=\sigma_i\left(W_iz_{i-1}+b_i\right),
\end{equation}
where $z_i\in\mathbb{R}^{n_{z_i}}$ is the output vector of layer $i$ with the features $v$ being the input to the first layer (i.e., $v=z_0$) and the labels $l$ being the output of the last layer. The function $\sigma_i\colon\mathbb{R}^{n_{z_i}}\rightarrow\mathbb{R}^{n_{z_i}}$ is the activation function. 
%To learn nonlinear maps it is fundamental that activation functions of the hidden layers, i.e., the layers between input and output layer, contain some nonlinearities.
Common nonlinear activation functions are the sigmoid function, hyperbolic tangent, or the ReLU function.\cite{Ding2018} The training of the ANN is achieved by iteratively adapting the weight matrices $W_i\in\mathbb{R}^{n_{z_i}\times n_{z_{i-1}}}$ and the bias vectors $b_i\in\mathbb{R}^{n_{z_i}}$, so that a user-defined loss function is minimized. For regression problems, one of the most commonly used loss functions is the mean squared error (MSE)
\begin{equation*}
    \mathrm{MSE}=\dfrac{1}{n_l}\sum_{i=1}^{n_l}(l_i-\hat{l}_i)^2,
\end{equation*}
where $\hat{l}_i$ are the values of the labels predicted by the ANN. Other loss functions supported by HILO-MPC include the mean absolute error (MAE) or the Huber loss. Optimization algorithms like gradient descent or stochastic gradient descent are used to minimize the loss. A number of minimization algorithms can be interfaced in HILO-MPC, e.g., the Adam algorithm, a modification of stochastic gradient descent. These minimization algorithms return a gradient vector of the loss with respect to the output $l$ of the ANN. By applying the backpropagation algorithm, this gradient vector can then be used to update the weight matrices $W_i$ and the bias vectors $b_i$ throughout the ANN.\cite{Rumelhart1986} The training is stopped when there is no significant change in the loss over a defined number of iterations or if the number of maximum iterations is reached. %At the moment, HILO-MPC supports feed-forward fully connected ANNs.
% \textcolor{red}{XXX add an explanation how the training/realization is done/how this is interfaced to TensorFlow or pytorchXXX}

In HILO-MPC, the training of the ANNs itself is performed by the respective machine learning libraries, i.e., currently TensorFlow or PyTorch. The specific parameters and characteristics for the training of the ANN (e.g., epochs, batch size, loss function, etc.) as well as the training data are automatically processed and forwarded to the chosen machine learning library. After the training of the ANN is completed, all relevant information is extracted, and an equivalent CasADi graph is created for later use in the model or model-based controllers and estimators.

The following example shows how to define an ANN consisting of two layers with 10 neurons each, using a sigmoid function as activation function, and how to train the resulting model
\begin{minted}[bgcolor=bg,breaklines]{python}
from hilo_mpc import ANN, Dense

# Initialize NN
ann = ANN(features, labels)
ann.add_layers(Dense(10, activation='sigmoid'))
ann.add_layers(Dense(10, activation='sigmoid'))
# Add training data set
ann.add_data_set(df)
# Set up NN
ann.setup()
# Train the NN
batch_size = 64
epochs = 1000
ann.train(batch_size, epochs, validation_split=.2, patience=100, scale_data=True)
\end{minted}
where \mintinline{python}{Dense(...)} creates a fully connected layer. One can simply replace specific variables of a model with the predictions of the ANN by the command \mintinline{python}{model.substitute_from(ann)}. The variables to be replaced will be automatically detected and exchanged with the corresponding predictions of the ANN. Alternatively, it is also possible to directly add the ANN to the model via summation or subtraction.

\subsubsection{Gaussian Processes}

\begin{figure}
    \centering
    \inputpdf{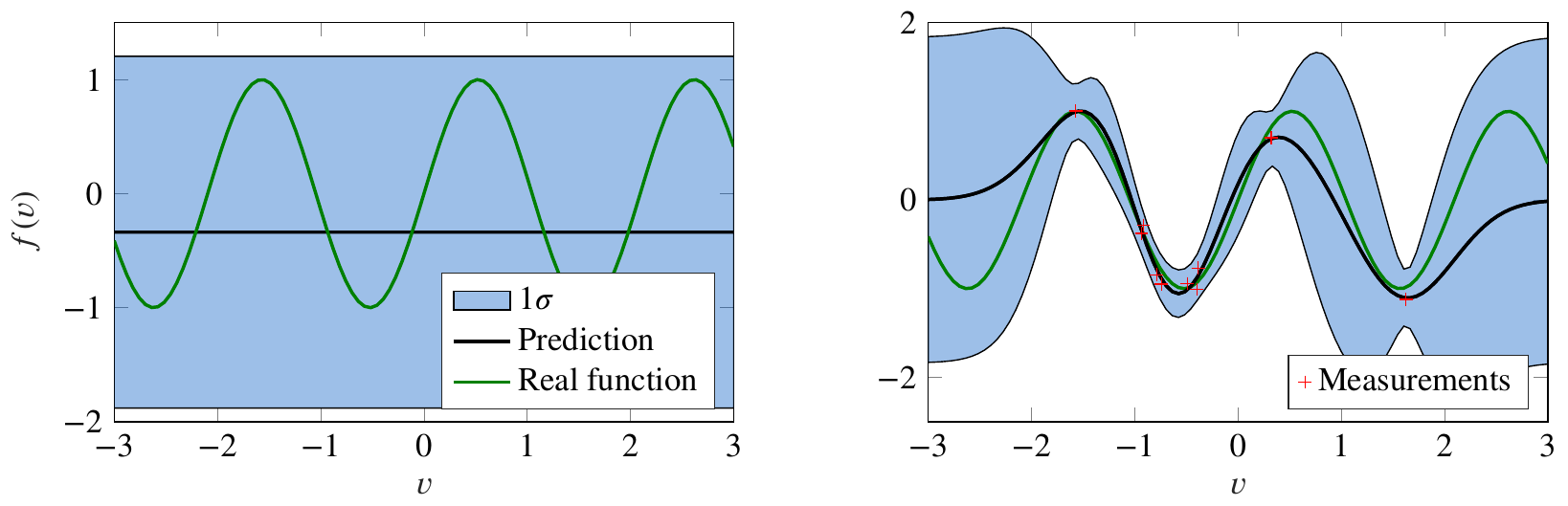}
    \caption{Simple example of a GP fitting to a sinusoidal function. The left figure shows a non-zero prior mean with $1\sigma$ standard deviation. The right figure shows the posterior of the GP conditioned on the measurements with optimal hyperparameters. The measurements are affected by a zero-mean white noise.}
    \label{fig:gp_training}
\end{figure}

A Gaussian process (GP) is a probability distribution over a function in which any finite subset of random variables has a joint Gaussian distribution.\cite{Rasmussen2006}
%, see e.g.\cite{Rasmussen2006}
Besides being less prone to overfitting, one key advantage of using a GP is that it naturally measures the output uncertainty by estimating the variance of the predictions. GPs can be used to approximate input-output mappings 
\begin{equation*}
    l=f(v)+\varepsilon,
\end{equation*}
where the output $l\in\mathbb{R}$ is affected by some zero-mean normally distributed measurement noise $\varepsilon\sim\mathcal{N}(0,\sigma_n^2)$ with variance $\sigma_n^2$.
Starting from a given prior distribution, described by a mean function and a covariance function, the GP is trained on observed data (see Figure~\ref{fig:gp_training}).

In general, both the prior mean function and the prior covariance function depend on so-called hyperparameters. During the training, the hyperparameters are obtained via optimization, i.e., the hyperparameters are chosen such that the resulting GP can be used to infer the input-output mapping given available training data. Typically, the prior mean function is assumed to be zero,\cite{Kocijan2003} but HILO-MPC also supports other prior mean functions. The covariance function, or kernel, is often assumed to be a smooth function that is infinitely differentiable. A number of covariance functions that meet this criterion are provided, e.g., the stationary squared exponential and rational quadratic kernels. The prior means, as well as the kernels, can be combined with each other through summation or multiplication.

The objective of the optimization problem that underlies the training step is usually to maximize the log marginal likelihood\cite{Rasmussen2006}
\begin{equation*}
    \log\left(p(l|V)\right)=-\dfrac{1}{2}l^\Tr\left(K+\sigma_n^2I\right)^{-1}l-\dfrac{1}{2}\log\left(\left|K+\sigma_n^2I\right|\right)-\dfrac{n}{2}\log\left(2\pi\right),
\end{equation*}
where $V\in\mathbb{R}^{n_v\times n_d}$ describes the matrix containing the $n_d$ training inputs, $K\in\mathbb{R}^{n_d\times n_d}$ is the covariance matrix resulting from the used kernel and the supplied training data, and $I$ is the identity matrix of corresponding size. The element $(i,j)$ of $K$ is $k(v_i,v_j)$, where $k\colon\mathbb{R}^{n_v}\times\mathbb{R}^{n_v}\rightarrow\mathbb{R}$ is the kernel. The predictive mean $\bar{f}$ and variance $\sigma^2$ of the trained GP can be obtained for test data $w$ as follows
\begin{align*}
    \bar{f}(w)&=\tilde{k}^\Tr(K+\sigma_n^2I)^{-1}l,\\
    \sigma^2(w)&=k(w,w)-\tilde{k}^\Tr(K+\sigma_n^2I)^{-1}\tilde{k},
\end{align*}
where the $i$-th element of $\tilde{k}\in\mathbb{R}^{n_d}$ is $k(v_i,w)$. The following example outlines how GPs can be setup
\begin{minted}[bgcolor=bg,breaklines]{python}
from hilo_mpc import GP, SquaredExponentialKernel

# Initialize kernel
kernel = SquaredExponentialKernel(variance=0.002)
# Initialize GP
gp = GP(features, labels, kernel=kernel)
# Add training data set
gp.set_training_data(train_in, train_out)
# Set up GP
gp.setup()
# Fit the GP
gp.fit_model()
\end{minted}
Similarly to ANNs, the GPs can be integrated in models.

\subsection{Control Module}
The control module realizes feedback controllers of proportional–integral–derivative (PID) type, linear quadratic regulators (LQR), as well as optimal and MPC controllers. The latter are outlined in the following.

\subsubsection{Model Predictive Control}

Model predictive control repeatedly solves a finite-horizon optimal control problem (OCP).\cite{grune2017nonlinear,rawlings2017model,findeisen2002introduction} Optimal control problems and MPC problems for continuous-time systems are realized in sampled-data fashion in HILO-MPC as outlined in the following.\cite{findeisen2003towards} Equivalent formulations for  discrete-time nonlinear system are also possible. At a sampling time $t_0$ the optimal control problem to solve reads
\begin{mini!}
    {u(\cdot)}{\int^{t_0+T}_{t_0} l(t,x(t),z(t),u(t),p) \mathrm{d}t + e(t_0+T,x(t_0+T),z(t_0+T),p),\label{subeq:obj1}}{\label{eq:problem-mpc}}{}
    \addConstraint{\dot{x}(t)}{=f(t,x(t),z(t),u(t),p),\quad\label{subeq:model1}}{x(t_0)=\hat{x}(t_0)}
    \addConstraint{0}{=q(t,x(t),z(t),u(t),p)}{}
    \addConstraint{0}{\geq g(t,x(t),z(t),u(t),p)\label{subeq:path-con}}{}
    \addConstraint{0}{\geq g_T(t_0+T,x(t_0+T),z(t_0+T),p)\label{subeq:terminal-con}}{}
    \addConstraint{y(t)}{=h(t, x(t),z(t),u(t),p)\label{subeq:meas-eq}}{}
    \addConstraint{u(t)}{=u(t_0 + T_c)\,\ \text{for} \,\ t\geq t_0 + T_c}{}  % looks better if we don't put the for-clause in the 3rd bracket pair
    \addConstraint{}{\text{for} \quad t \in [t_0, t_0+T] \subset \mathbb{R}.}{}
\end{mini!}
% \begin{subequations}
%     \label{eq:problem-mpc}
%     \begin{alignat}{2}
%         &\!\min_{u(\cdot)}&\qquad& \int^{t_0+T}_{t_0} l(t,x(t),z(t),u(t),p) \mathrm{d}t + e(t_0+T,x(t_0+T),z(t_0+T),p),\label{subeq:obj1}\\
%         &\text{s.t.}&&\dot{x}(t)=f(t,x(t),z(t),u(t),p),\quad x(t_0)=\hat{x}(t_0),\label{subeq:model1}\\
%         &&& 0=q(t,x(t),z(t),u(t),p), \\
%         &&&g(t,x(t),z(t),u(t),p) \leq 0, \label{subeq:path-con}\\
%         &&&g_T(t_0+T,x(t_0+T),z(t_0+T),p) \leq 0,\label{subeq:terminal-con}\\
%         &&& y(t) = h(t, x(t),z(t),u(t),p),\label{subeq:meas-eq}\\
%         &&& u(t) = u(t_0 + T_c), \,\ \text{for} \,\ t\geq t_0 + T_c, \\
%         &&& \text{for} \quad t \in [t_0, t_0+T] \subset \mathbb{R}.
%     \end{alignat}
% \end{subequations} 

Here, $u(\cdot)$ is the optimal input function, $l:\mathbb{R}\times\mathbb{R}^{n_x}\times\mathbb{R}^{n_z}\times\mathbb{R}^{n_u} \times \mathbb{R}^{n_p} \rightarrow \mathbb{R}$ is the stage cost, $e:\mathbb{R}\times\mathbb{R}^{n_x}\times\mathbb{R}^{n_z} \times \mathbb{R}^{n_p} \rightarrow \mathbb{R}$ the terminal cost, $T$ is the prediction horizon, and $T_c$ the control horizon. The vector $\hat{x}(t_0)$ contains the measured or estimated states at the current sampling time $t_0$ and $t$ is the time variable. The equation $g:\mathbb{R}\times\mathbb{R}^{n_x}\times\mathbb{R}^{n_z}\times\mathbb{R}^{n_u} \times \mathbb{R}^{n_p} \rightarrow \mathbb{R}^{n_g}$ represents path constraints and  $g_T:\mathbb{R}\times\mathbb{R}^{n_x}\times\mathbb{R}^{n_z} \times \mathbb{R}^{n_p} \rightarrow \mathbb{R}^{n_{g,\text{T}}}$ are terminal constraints.
% \begin{remark}
% Any of the equations in \eqref{eq:problem-mpc} could be at least partially learned from data (see Figure~ \ref{fig:scheme}). For example in \cite{matschek2019learning} the output function \label{subeq:meas-eq} is learned, in \cite{rosolia2017autonomous,rosolia2017learning,brunner2017repetitive} the terminal constraints \eqref{subeq:terminal-con} and in \cite{bujarbaruah2020learning,Leopoldo2017} the path constraints. Also the objective function can be learned, see e.g. \cite{tamar2017learning,BECKENBACH201860,Bradford2018}. HILO-MPC allows to use machine learning models in any of these components.
% \end{remark}
\begin{remark}
In principle any of the equations in \eqref{eq:problem-mpc} could be completely or partially learned from data (see Figure~ \ref{fig:scheme}), using the learning approaches outlined in Section \ref{sec:learning}. Table \ref{tab:what_is_learned} gives an overview of other works where different components are learned. These cases can be handled using the \emph{machine learning} module.
\end{remark}

\begin{table*}[htb]
    \caption{Selected, non-extensive overview of learned components of MPC and optimal control problems in the literature.}
    \label{tab:what_is_learned}
    \centering
    \begin{tabular}{ll}
        \toprule
        What is learned? & Related work \\
        \midrule
        reference/desired output function \eqref{subeq:meas-eq} & Matschek et al.\cite{matschek2019learning}, Carron et al. \cite{carron2019data}, \dots \\
        terminal constraints \eqref{subeq:terminal-con} & Rosolia et al.\cite{rosolia2017autonomous}, Rosolia and Borrelli\cite{rosolia2017learning}, Brunner et al.\cite{brunner2017repetitive}, \dots \\
        path constraints \eqref{subeq:path-con} & Bujarbaruah et al.\cite{bujarbaruah2020learning}, Armesto et al.\cite{Leopoldo2017}, Holzmann et al. \cite{holzmann2022} , \dots\\
        objective function \eqref{subeq:obj1} & Tamar et al.\cite{tamar2017learning}, Beckenbach et al.\cite{BECKENBACH201860}, Bradford and Imsland\cite{Bradford2018}, \dots \\
        \bottomrule
    \end{tabular}
\end{table*}

Solving Problem \eqref{eq:problem-mpc} directly is challenging, since $u(\cdot)$ and the constraints are infinite-dimensional. HILO-MPC uses \emph{direct approaches} to transform the problem into an equivalent finite-dimensional optimization problem. These approaches parameterize the input with a finite number of parameters, for example, using piece-wise constant inputs, and enforce the constraints only on a finite number of points. The resulting problem becomes
\begin{mini!}
    {\mathbf{u}}{\sum_{i = k}^{N}  \int_{t_i}^{t_{i} + \Delta t} l(t,x(t),z(t),u_i,p) \mathrm{d}t + e(t_k+T,x(t_k+T),z(t_k+T),p),}{\label{eq:disc-mpc}}{}
    \addConstraint{x(t_{i+1})}{= x(t_{i}) + \int_{t_i}^{t_{i} + \Delta t} f(t,x(t),z(t),u_i,p) \mathrm{d}t}{}
    \addConstraint{x(t_k)}{=\hat{x}(t_k)}{}
    \addConstraint{0}{=q(t_i,x(t_i),z(t_i),u_i,p)}{}
    \addConstraint{0}{\geq g(t_i,x(t_i),z(t_i),u_i,p)}{}
    \addConstraint{0}{\geq g_T(t_k+T,x(t_k+T),z(t_k+T),p)}{}
    \addConstraint{y(t_i)}{=h(t_i, x(t_i),z(t_i),u_i,p)}{}
    \addConstraint{u_i}{=u_{N_c},\,\ \text{for} \,\ i\geq N_c}{}  % looks better if we don't put the for-clause in the 3rd bracket pair
    \addConstraint{}{\text{for} \,\ i \in [k, N] \subset \mathbb{N}_0,}{}
\end{mini!}
% \begin{subequations}
% \label{eq:disc-mpc}
%     \begin{alignat}{2}
%         &\!\min_{\mathbf{u}}&\qquad& \sum_{i = k}^{N}  \int_{t_i}^{t_{i} + \Delta t} l(t,x(t),z(t),u_i,p) dt + e(t_k+T,x(t_k+T),z(t_k+T),p),\\
%         &\text{s.t.}&& x(t_{i+1})= x(t_{i}) + \int_{t_i}^{t_{i} + \Delta t} f(t,x(t),z(t),u_i,p) dt,\\ 
%         &&& x(t_k)=\hat{x}(t_k),\\
%         &&& 0=q(t_i,x(t_i),z(t_i),u_i,p), \\
%         &&&g(t_i,x(t_i),z(t_i),u_i,p) \leq 0, \\
%         &&&g_T(t_k+T,x(t_k+T),z(t_k+T),p) \leq 0, \\
%         &&& y(t_i) = h(t_i, x(t_i),z(t_i),u_i,p),\\
%         &&& u_i = u_{N_c}, \,\ \text{for} \,\ i\geq N_c, \\
%         &&& \text{for} \quad i \in [k, N] \subset \mathbb{N}_0,
%     \end{alignat}
% \end{subequations} 
where $\Delta t$ is the sampling time, $N_c = \text{ceil}(T_c/\Delta t)$ is the control horizon and $\mathbf{u}=\begin{bmatrix}u_k&\dots&u_{N_c}\end{bmatrix}$ is the sequence of piece-wise constant inputs applied to the plant, i.e., $u(t) = u_i, \,\ \forall t \in [t_i,\,t_i + \Delta t) \subset \mathbb{R}$. The function $\text{ceil}\colon\mathbb{R}\rightarrow\mathbb{Z}$ is the ceiling function, i.e., a function that returns the smallest integer that is larger than a given real number.
HILO-MPC also implements the multiple shooting approach.\cite{BOCK19841603} The integration of the system model in between the shooting points can be done with Runge-Kutta methods of various order, orthogonal collocation,\cite{Oh1977} or the dynamics can be integrated with CVODES or IDAS solvers.\cite{Hindmarsh2005} The default method is orthogonal collocation, while any other method can be selected if needed. Note that for simplicity piece-wise, constant inputs are assumed at every control interval, which can be generalized to introducing virtual inputs. 
Hence, the real formulation of Problem \eqref{eq:problem-mpc} depends on which method is used for the discretization. Here, no further details of the formulation of the approximated problem are provided. See the cited papers for more information.
\begin{remark}
Note that it is also possible to optimize the sampling intervals. This will add to Problem \eqref{eq:disc-mpc} the vector $\begin{bmatrix}\Delta t_k&\dots&\Delta t_{N}\end{bmatrix}$ as optimization variable. This allows the solution of minimum time problems, and optimal sampling time problems. This frequently appears for example in network controlled systems.\cite{heemels2010networked,lucia2016predictive}
\end{remark}

The stage cost $l(\cdot)$ and arrival cost $e(\cdot)$ depend on the type of optimal control formulation/MPC formulation implemented. In the following we sketch different MPC formulations contained in the toolbox. For simplicity of explanation, we focus on quadratic cost functions; however, HILO-MPC is not limited to those.

\paragraph{MPC for set point tracking}

In set point tracking MPC\cite{Matschek2019} the variables $x$, $u$ and $y$ are desired to track a known reference, which can be formulated in terms of the cost function and terminal cost, for example, as
\begin{align*}
    l(x(t),u(t),y(t)) &= \Vert x(t) - x_r\Vert^2_Q + \Vert u(t) - u_r\Vert^2_R +  \Vert y(t) - y_r\Vert^2_P, \\
    e(x(t_0+T),y(t_0+T)) &= \Vert x(t_0+T) - x_r\Vert^2_{Q_{\text{T}}} +  \Vert y(t_0+T) - y_r\Vert^2_{P_{\text{T}}}.
\end{align*}
Here $x_r$, $u_r$, and $y_r$ are fixed references and $Q \in \mathbb{R}^{n_x \times n_x}$, $R \in \mathbb{R}^{n_u \times n_u}$, $P \in \mathbb{R}^{n_y \times n_y}$, $P_{\text{T}} \in \mathbb{R}^{n_y \times n_y}$, and $ Q_{\text{T}} \in \mathbb{R}^{n_x \times n_x}$ are weighting matrices. 
To outline the simplicity to formulate MPC problems, consider the one-track car model as discussed in section \ref{sec:modeling_module}. The goal is that the car should track a set point reference speed  $v_{\text{r}} = 2 \,\ \mathrm{m}/\mathrm{s}$. We consider a  prediction horizon of 20 steps. The corresponding MPC controller can be formulated as
\begin{minted}[bgcolor=bg,breaklines]{python}
from hilo_mpc import NMPC

nmpc = NMPC(model)
nmpc.horizon = 20
nmpc.quad_stage_cost.add_states(names=['v'], ref=2, weights=10)
nmpc.quad_term_cost.add_states(names=['v'], ref=2, weights=10)
nmpc.setup()
\end{minted}
The resulting MPC controller can be easily used in a control loop
\begin{minted}[bgcolor=bg,breaklines]{python}
# Choose number of steps
n_steps = 100

# Begin simulation loop
for step in range(n_steps):
    # find optimal input
    u = nmpc.optimize(x0)
    # Simulate the plant
    model.simulate(u=u)
    # Get the current states
    x0 = sol['xf']
\end{minted}
The solution of the MPC (e.g., input and state sequences) are stored in the \mintinline{python}{mpc.solution()} object.

\paragraph{MPC for trajectory-tracking}

For trajectory tracking problems the variables need to track a time-varying reference.\cite{Matschek2019} Considering a quadratic cost function, this can be formulated as
\begin{align*}
\label{eq:mpc-traj-tracking}
    l(x(t),u(t),y(t)) &= \Vert x(t) - x_r(t)\Vert^2_Q + \Vert u(t) - u_r(t)\Vert^2_R +  \Vert y(t) - y_r(t_i)\Vert^2_P, \\
    e(x(t_0+T),y(t_0+T)) &= \Vert x(t_0+T) - x_r(t_0+T) \Vert^2_{Q_{\text{T}}} +  \Vert y(t_0+T) - y_r(t_0+T) \Vert^2_{P_{\text{T}}}.
\end{align*}
Here $x_r : \mathbb{R} \rightarrow \mathbb{R}^{n_x}$, $u_r: \mathbb{R} \rightarrow \mathbb{R}^{n_u}$, and $y_r: \mathbb{R} \rightarrow \mathbb{R}^{n_y}$ are time-varying references. For example, a trajectory tracking problem for a single-track car model can be defined by
\begin{minted}[bgcolor=bg,breaklines]{python}
from hilo_mpc import NMPC

nmpc = NMPC(model)
nmpc.horizon = 20

t = nmpc.create_time_variable()
traj_x = 30 - 14 * cos(t)
traj_y = 30 - 16 * sin(t)

nmpc.quad_stage_cost.add_states(names=['px', 'py'], ref=[traj_x, traj_y], weights=[10, 10], trajectory_tracking=True)
nmpc.quad_term_cost.add_states(names=['px', 'py'], ref=[traj_x, traj_y], weights=[10, 10], trajectory_tracking=True)

nmpc.setup()
\end{minted}

\paragraph{MPC for path following}

While in the MPC for trajectory tracking both the \emph{value of the reference and time are fixed simultaneously}, MPC for path following  exploits additional degrees of freedom by choosing \emph{when} to be where on the path.\cite{Matschek2019,faulwasser2016} Do do so, we augment the model with a \emph{virtual path state}
\begin{mini!}
    {u(\cdot),u_{\text{pf}}(\cdot)}{\int_{t_0}^{t_0+T} l(t,x(t),z(t),u(t),p) \mathrm{d}t + e(t_0+T,x(t_0+T),z(t_0+T),p),}{}{}
    \addConstraint{\dot{x}(t)}{=f(t, x(t),z(t),u(t),p),\quad}{x(t_0)=\hat{x}(t_0)}
    \addConstraint{0}{=q(t, x(t),z(t),u(t),p)}{}
    \addConstraint{\dot{\theta}}{=u_{\text{pf}},\quad}{\theta(t_0)=0}
    \addConstraint{0}{\geq g(t,x(t),z(t),u(t),p, \epsilon)}{}
    \addConstraint{0}{\geq g_T(t_0+T,x(t_0+T),z(t_0+T),p)}{}
    \addConstraint{y(t)}{=h(t, x(t), z(t), u(t), p)}{}
    \addConstraint{}{\text{for} \,\ t \in [t_0, t_0+T] \subset \mathbb{R}.}{}
\end{mini!}
% \begin{subequations}
%     \begin{alignat}{2}
%         &\!\min_{u(\cdot),u_{\text{pf}}(\cdot)}&\qquad& \int_{t_0}^{t_0+T} l(t,x(t),z(t),u(t),p) \mathrm{d}t + e(t_0+T,x(t_0+T),z(t_0+T),p),\\
%         &\text{s.t.}&&\dot{x}(t)=f(t, x(t),z(t),u(t),p),\quad x(t_0)=\hat{x}(t_0), \label{subeq:model1}\\
%         &&& 0=q(t, x(t),z(t),u(t),p), \\
%         &&& \dot{\theta}= u_{\text{pf}},\quad \theta(t_0)=0, \\
%         &&&g(t,x(t),z(t),u(t),p, \epsilon) \leq 0, \\
%         &&&g_T(t_0+T,x(t_0+T),z(t_0+T),p) \leq 0,\\
%         &&& y(t) = h(t, x(t), z(t), u(t), p), \\
%         &&& \text{for} \quad t \in [t_0, t_0+T] \subset \mathbb{R}, 
%     \end{alignat}
% \end{subequations} 
Here $\theta \in \mathbb{R}^{n_\theta}$ is a virtual path state vector and $u_{\text{pf}} \in \mathbb{R}^{n_{u,\theta}}$ is the virtual input that the controller can choose. Hence, the objective function becomes
\begin{align*}
    l(x(t),u(t),y(t)) &= \Vert x(t) - x_r(\theta(t))\Vert^2_Q + \Vert u(t) - u_r(\theta(t))\Vert^2_R +  \Vert y(t) - y_r(\theta(t))\Vert^2_P, \\
    e(x(t_0+T),y(t_0+T)) &= \Vert x(t_0+T) - x_r(\theta(t_0+T)) \Vert^2_{Q_{\text{T}}} +  \Vert y(t_0+T) - y_r(\theta(t_0+T)) \Vert^2_{P_{\text{T}}}.
\end{align*}
To force the controller to move only in the direction of the path usually a lower bound on $u_{\text{pf}}$ is added, i.e., $u_{\text{pf}}>u_{\text{pf,min}}$ with $u_{\text{pf,min}} \in \mathbb{R}^{n_{u,\theta}}_{+}$.

It is also possible to track a constant $u_{\text{pf}}$, so that the MPC tries to maintain a constant \emph{speed} of the virtual state
\begin{align*}
    l(x(t),u(t),y(t)) &= \Vert x(t) - x_r(\theta(t))\Vert^2_Q + \Vert u(t_i) - u_r(\theta(t))\Vert^2_R +  \Vert y(t) - y_r(\theta)\Vert^2_P + \Vert u_{\text{pf}} - u_{\text{pf,ref}}\Vert_{R_{\text{pf}}}^2,  \\
    e(x(t_0+T),y(t_0+T)) &= \Vert x(t_0+T) - x_r(\theta(t_0+T))\Vert^2_Q +  \Vert y(t_0+T) - y_r(\theta(t_0+T))\Vert^2_P.
\end{align*}

In comparison to other toolboxes, path following MPC/optimal control problems can be easily formulated. The user needs to activate the path following mode for the desired variables, as shown in the following
\begin{minted}[bgcolor=bg,breaklines]{python}
from hilo_mpc import NMPC

nmpc = NMPC(model)
nmpc.horizon = 20

theta = nmpc.create_path_variable()
traj_x = 30 - 14 * cos(theta)
traj_y = 30 - 16 * sin(theta)

nmpc.quad_stage_cost.add_states(names=['px', 'py'], ref=[traj_x, traj_y], weights=[10, 10], path_following=True)
nmpc.quad_term_cost.add_states(names=['px', 'py'], ref=[traj_x, traj_y], weights=[10, 10], path_following=True)

nmpc.setup()
\end{minted}
Note that the toolbox allows to mix the previous problem formulations with minimum effort, e.g., the user can track some constant references for some of the variables, while performing path tracking for the other variables.

\paragraph{Soft constraints}

Soft constraints are typically used in optimal control/MPC to avoid infeasibility, conservative solutions due to active constraints, or to speed up computations.\cite{kerrigan2000soft,richards2015fast} When soft constraints are selected, slack variables $\epsilon_{\text{p}} \in \mathbb{R}^{n_g}$ and $\epsilon_{\text{T}} \in \mathbb{R}^{n_{g,\text{T}}} $ are automatically added to the path and terminal constraints, respectively, as follows
\begin{mini!}
    {u(\cdot), \epsilon_{\text{T}}, \epsilon_{\text{p}}}{\int_{t_0}^{t_0+T} l(\cdot) \mathrm{d}t + e(\cdot) + \Vert \epsilon_{\text{s}} \Vert^2_{E_{\text{p}}} + \Vert \epsilon_{\text{T}} \Vert^2_{E_{\text{T}}},}{\label{eq:problem-mpc-soft}}{}
    \addConstraint{\dot{x}(t)}{=f(t,x(t),z(t),u(t), p),\quad}{x(0)=\hat{x}_0}
    \addConstraint{0}{=q(t, x(t),z(t),u(t), p)}{}
    \addConstraint{\epsilon_{\text{p}}}{\geq g(t,x(t),z(t),u(t), p)}{}
    \addConstraint{\epsilon_{\text{T}}}{\geq g_T(t_0+T,x(t_0+T),z(t_0+T),p)}{}
    \addConstraint{y(t)}{=h(t, x(t),z(t), u(t),p)}{}
    \addConstraint{0}{\leq\epsilon_{\text{p}} \leq \epsilon_{\text{p,max}}, \quad 0 \leq \epsilon_{\text{T}} \leq \epsilon_{\text{T, max}}}{}
    \addConstraint{}{\text{for} \,\ t \in [t_0, t_0+T] \subset \mathbb{R},}{}
\end{mini!}
% \begin{subequations}
%     \label{eq:problem-mpc-soft}
%     \begin{alignat}{2}
%         &\!\min_{u(\cdot), \epsilon_{\text{T}}, \epsilon_{\text{p}} }&\qquad& \int_{t_0}^{t_0+T} l(\cdot) \mathrm{d}t + e(\cdot) + \Vert \epsilon_{\text{s}} \Vert^2_{E_{\text{p}}} + \Vert \epsilon_{\text{T}} \Vert^2_{E_{\text{T}}}, \\
%         &\text{s.t.}&&\dot{x}(t)=f(t,x(t),z(t),u(t), p),\quad x(0)=\hat{x}_0, \\
%         &&& 0 = q(t, x(t),z(t),u(t), p),\\
%         &&&g(t,x(t),z(t),u(t), p) \leq \epsilon_{\text{p}}, \\
%         &&&g_T(t_0+T,x(t_0+T),z(t_0+T),p) \leq \epsilon_{\text{T}},\\
%         &&& y(t) = h(t, x(t),z(t), u(t),p), \\
%         &&& 0 \leq \epsilon_{\text{p}} \leq \epsilon_{\text{p,max}}, \quad 0 \leq \epsilon_{\text{T}} \leq \epsilon_{\text{T, max}},\\
%         &&& \text{for} \quad t \in [t_0, t_0+T] \subset \mathbb{R},
%     \end{alignat}
% \end{subequations}
where and $E_{\text{p}}\in \mathbb{R}^{n_g \times n_{g}}$ and $E_{\text{T}}\in \mathbb{R}^{n_{g,\text{T}} \times n_{g,\text{T}}}$ are  weighting matrices that limit the increase of the slack variables and can be chosen by the user, and $\epsilon_{\text{p,max}}$ and $\epsilon_{\text{T,max}}$ are the maximum constraint violations of path and terminal constraints, respectively.

\subsection{Estimator Module}
Estimating states and parameters from measurements is indispensable in control if not all variables can be accessed directly. The estimator module provides access to a moving horizon estimator, Kalman filters, and particle filters.

\subsubsection{Moving Horizon Estimation}

Moving horizon estimators (MHE) obtain state estimates based on the solution of an optimization problem similar to the MPC problem.\cite{rawlings2017model,johansen2011introduction,findeisen2003state} We focus on moving horizon estimators operating on equidistant sampling times. At these sampling times measurements are taken. Due to the discrete measurements, the objective function in an MHE is usually given in discrete form. For simplicity we indicate with $(\cdot)_{i|k}$ the variable at time $t_i$, i.e., for the MHE problem solved at time $t_k$ we obtain
\begin{mini!}
    {x_{k-N|k},p_k, w_{(\cdot)|k},z_{(\cdot)|k}}{\left\Vert \begin{bmatrix} x_{k-\bar{N}|k} - \hat{x}_{k-N|k} \\ p_{k} - \hat{p}_{k} \end{bmatrix} \right\Vert^2_{P_k} +  \sum^k_{i=k-\bar N}  \Vert \hat{y}_i - y_{i|k} \Vert^2_R + \Vert w_{i|k}\Vert^2_W,}{\label{eq:problem-mhe}}{}
    \addConstraint{x_{i+1|k}}{=x_{i|k} + \int_{t_i}^{t_i + \Delta t} \left( f(t,x(t),z(t),\hat{u}(t), p_k) \right) \mathrm{d}t + w_{i|k} \label{subeq:mhe-model1}}{}
    \addConstraint{y_{i|k}}{=h(t_i,x_{i|k},z_{i|k},\hat{u}_i,p_k) + v_{i|k}}{}
    \addConstraint{0}{\geq g(t_i,x_{i|k},u_i)}{}
    \addConstraint{}{\text{for} \,\ i \in [k-\bar{N}, k], \,\ k,\bar{N} \in \mathbb{N},}{}
\end{mini!}
% \begin{subequations}
%     \label{eq:problem-mhe}
%     \begin{alignat}{2}
%         &\!\min_{x_{k-N|k},p_k, w_{(\cdot)|k},z_{(\cdot)|k}}&\qquad& \left\Vert \begin{bmatrix} x_{k-\bar{N}|k} - \hat{x}_{k-N|k} \\ p_{k} - \hat{p}_{k} \end{bmatrix} \right\Vert^2_{P_k} +  \sum^k_{i=k-N}  \Vert \hat{y}_i - y_{i|k} \Vert^2_R + \Vert w_{i|k}\Vert^2_W, \\
%         &\text{s.t.}&&x_{i+1|k}= x_{i|k} + \int_{t_i}^{t_i + \Delta t} \left( f(t,x(t),z(t),\hat{u}(t), p_k) + w(t) \right) dt, \label{subeq:mhe-model1}\\
%         &&& y_{i|k} = h(t_i,x_{i|k},z_{i|k},\hat{u}_i,p_k) + v_{i|k},\\
%         &&&g(t_i,x_{i|k},u_i) \leq 0, \\
%         &&& \text{for} \,\ i \in [k-\bar{N}, k], \,\ k,\bar{N} \in \mathbb{N},
%     \end{alignat}
% \end{subequations} 
where $\bar{N}$ is the horizon length, $\hat{y}_i$ the output measurements, and $\hat{u}_i$ the input measurements at time $t_i$. The first term of the objective function is the \emph{arrival cost}, while the second and third weigh the measurement and state noise, respectively.\cite{rawlings2017model,allgower1999nonlinear} The matrices $R \in \mathbb{R}^{n_y \times n_y}$, $W \in \mathbb{R}^{n_x \times n_x}$ and $P_k \in \mathbb{R}^{(n_x+n_p) \times (n_x + n_p)}$ are the weighting matrices for the outputs, state noise, and arrival cost. The optimization variables are: the state $x_{k-\bar{N}|k}$, i.e., the state at the beginning of the horizon, the state noise $w_{(\cdot)|k} = \{ w_{i|k}, \,\ \forall	 i \in [k-\bar{N},k] \}$, the algebraic states (for DAE systems) $z_{(\cdot)|k} = \{ z_{i|k}, \,\ \forall	 i \in [k-\bar{N},k] \}$ and the system parameters $p_k$. Note that the parameters are considered constant in the horizon, but can be updated every time the optimization is run, to adapt to the new measurements.
%Similar to the optimal control/MPC problem direct solution approaches to solve \eqref{eq:problem-mhe} as for the MPC case. 
For the single-track car model, an MHE with $R=W=P_{k}=10 I$ (where $I$ is the identity matrix of appropriate dimensions) can be easily defined as follows
\begin{minted}[bgcolor=bg,breaklines]{python}
from hilo_mpc import MHE

mhe = MHE(model)
mhe.horizon = 20
mhe.quad_arrival_cost.add_states(weights=[10,10,10,10], guess=x0_est)
mhe.quad_stage_cost.add_measurements(weights=[10,10,10,10])
mhe.quad_stage_cost.add_state_noise(weights=[10,10,10,10]])
mhe.setup()
\end{minted}
After this setup, it can be deployed for estimation simply by
\begin{minted}[bgcolor=bg,breaklines]{python}
# Set number of steps
n_steps = 100

# Simulation/estimation loop
for step in range(n_steps):
    # Simulate plant with input
    model.simulate(u=u)
    # Get measurements
    y_meas = model.solution['y'][:, -2]
    # Pass the measurements to the MHE
    mhe.add_measurements(y_meas, u_meas=u)
    # Estimate states 
    x0 =  mhe.estimate()
\end{minted}
Similar to other objects, the solution of the MHE is stored in \mintinline{python}{mhe.solution()}.

\subsubsection{Kalman Filters and Particle Filters}

Besides an MHE, Kalman filters (KFs) and particle filters (PFs) are provided for estimation. KFs allow for the estimation of observable states via available measurement data. Generally, a KF consists of two steps: the prediction step and the update step. In the prediction step, the estimated states from the previous iteration are propagated through the model dynamics to obtain preliminary values for the states at the current time step, the so-called \emph{a priori} estimates. These \emph{a priori} estimates are then updated in the update step using the measurement data to obtain \emph{a posteriori} estimates. 

HILO-MPC contains KFs in various formulations. An unscented KF\cite{Wan2000} can simply be integrated by% in HILO-MPC can be easily set up as
\begin{minted}[bgcolor=bg,breaklines]{python}
from hilo_mpc import UKF

ukf = UKF(model)
ukf.setup()
\end{minted}

In a similar fashion  particle filters can be formulated and used.

\subsection{Embedded Module}
\label{sec:emb-mod}
Implementing optimization-based controllers/MPC controllers on embedded systems requires a tailored implementation. By now, a series of tailored automatic code generation tools for MPC exist.\cite{Houska2011a,Verschueren2019,zometa2013,domahidi2014forces}

HILO-MPC provides an interface for the code generation of MPC for linear time-invariant discrete-time model
predictive control, tailored for small embedded systems,
like microcontrollers.\cite{zometa2013}

The code generation is limited to discrete-time linear time-invariant systems of the form 
\begin{equation}
  \label{eqn:system}
  x^+ = Ax + Bu 
\end{equation}
subject to input and state constraints
%\begin{equation*}
  $\underline{u} \leq u \leq \overline{u}$, $\underline{x} \leq x \leq \overline{x}$,
%\end{equation*}
where $x$ and $u$ denote the state and input at the current sampling time,
and $x^+$ denotes the state at the next sampling time.

The considered linear MPC problem formulation  is
\begin{mini}
    {\mathbf{u}}{\frac{1}{2} \sum\limits_{j=0}^{N-1} (\|x_j\|^2_Q + \|u_j\|^2_R) 
    + \frac{1}{2} \|x_N\|^2_P,}{\label{eqn:mpcproblem}}{}
    \addConstraint{x_{j+1}}{= A x_j + B u_j,\qquad}{j = 0,\dots, N-1}
    \addConstraint{\underline{u}}{\leq u_j \leq \overline{u},}{j = 0,\dots, N-1}
    \addConstraint{\underline{x}}{\leq x_j \leq \overline{x},}{j = 0,\dots, N-1}
    \addConstraint{x_{0}}{=x,}{}
\end{mini}
% \begin{equation}
%   \label{eqn:mpcproblem}
%   \begin{aligned}
%     \underset{\mathbf{u}}{\text{minimize}}
%     & \;\; \frac{1}{2} \sum\limits_{j=0}^{N-1} (\|x_j\|^2_Q + \|u_j\|^2_R) 
%     + \frac{1}{2} \|x_N\|^2_P \\
%     \text{subject to}
%     & \;\; x_{j+1} = A x_j + B u_j, \;\; & j = 0,..., N-1,\\
%     & \;\; \underline{u} \leq u_j \leq \overline{u}, \;\; & j = 0,..., N-1, \\
%     & \;\; \underline{x} \leq x \leq \overline{x}, \;\; & j = 0,..., N-1,\\
% %    & \;\; \underline{f} \leq F x_N \leq \overline{f}, \\
%     & \;\; x_{0}=x
%   \end{aligned}
% \end{equation}

where the integer $ N \geq 2$ is the prediction horizon, the matrices $Q \inRpow{n_x
\times n_x}$, $R \inRpow{n_u \times n_u}$ and $P \inRpow{n_x \times n_x}$ are the state,
input, and terminal weighting matrix, respectively. The input sequence 
$\vb{u} = \begin{bmatrix}u_{0}&\dots&u_{N-1}\end{bmatrix}$ is the optimization variable. 
%We define the set $\mathcal{U} = \{ \vb{u} \; | \; \underline{u} \leq u_j \leq \overline{u}, \; j\!=\!0,...,N-1\}$
%, or  equivalently $\mathcal{U}=\{ \vb{u} \; | \; \underline{\vb{u}} \leq \vb{u} \leq \overline{\vb{u}}\}$ 
%and assume it contains the origin.
%Furthermore, mixed constraints are defined by the matrices $K_{x} \inRpow{q\times n}$ and
%$K_{u} \inRpow{q\times m}$, and the lower and upper bounds $\underline{e}$ and $\overline{e}$,
%respectively. Finally, the terminal state also needs to satisfy some box constraints, with
%$F \inRpow{r\times n}$ and the lower and upper bounds $\underline{f}$ and $\overline{f}$. 
For automatic code generation of the problem \eqref{eqn:mpcproblem} tailored towards microcontrollers, we rely  on \muaompc,\cite{zometa2013} a code generation tool 
that uses an augmented Lagrangian and Nesterov's fast gradient method.\cite{zometa2012, koegel2011}
By design, the algorithm's implementation is simple 
and relies only on matrix-vector operations (i.e., no divisions),
it can easily be warm started and computes good  MPC inputs
in very few iterations of the algorithm.
For well-behaved problems, the computations can be made more efficient by using 
fixed-point arithmetic (instead of
floating-point).

The code generation reformulates the MPC optimization problem (\ref{eqn:mpcproblem}) 
as a condensed parametric QP $\mathcal{P}(x)$
\begin{mini}
    {\vb{u} \in \mathcal{U}}{\frac{1}{2} \vb{u}^\Tr H \vb{u} + \vb{u}^\Tr g(x),}{\label{eqn:qp}}{}
    \addConstraint{\underline{\vb{z}}(x)}{\leq E \vb{u} \leq \overline{\vb{z}}(x),}{}
\end{mini}
% \begin{equation}
%   \label{eqn:qp}
%   \begin{aligned}
%     \underset{\vb{u} \in \mathcal{U}}{\text{minimize}}
%     & \;\; \frac{1}{2} \vb{u}^\Tr H \vb{u} + \vb{u}^\Tr g(x) \\
%     \text{subject to}
%     & \;\; \underline{\vb{z}}(x) \leq E \vb{u} \leq \overline{\vb{z}}(x),
%   \end{aligned}
% \end{equation}
where the constant Hessian matrix $H^\Tr\!=\!H > 0 \inRpow{Nn_u \times Nn_u}$ and the gradient vector $g(x)$
depend on the system (\ref{eqn:system}) and the objective function
in (\ref{eqn:mpcproblem}). Additionally, $g(x)$ depends on the current state $x$ (i.e., the parameter). % and the current reference trajectory.  
The set $\mathcal{U} = \{ \vb{u} \; | \; \underline{u} \leq u_j \leq \overline{u}, \; j\!=\!0,...,N-1\}$ defines the
input constraints.
The constant matrix $E$ and the constraint vectors $\underline{\vb{z}}(x)$ and
$\overline{\vb{z}}(x)$
depend on the 
constraints in (\ref{eqn:mpcproblem}). 

The generated code is highly portable C-code.
It is split into the data for the QP $\mathcal{P}(x)$ (e.g., the constant matrix $H$ 
and the parametric vector $g(x)$)
and the implementation of the optimization
algorithm to solve the QP.\cite{zometa2012, koegel2011}

\begin{figure*}
  \centering
  \begin{subfigure}{0.45\textwidth}
  \inputpdf{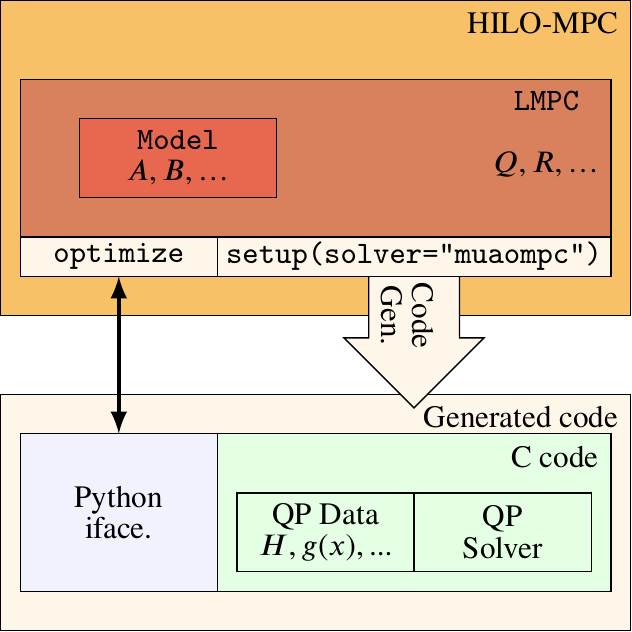}
  %\caption{Scheme of the embedded module.}

  \end{subfigure}
\begin{subfigure}{0.45\textwidth}
  \inputpdf{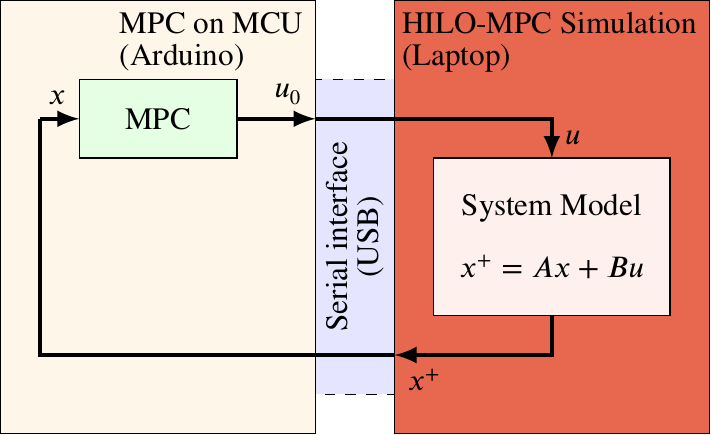}
  %\caption{Scheme of the considered hardware-in-the-loop simulation}
  %\label{fig:hil-sim}
  \end{subfigure}
\caption{A scheme of the Embedded module. Left, the typical used of the module, including automatic code generation provided by \muaompc. Right, a hardware-in-the-loop simulation used as example.}
\label{fig:hilo-muaompc} 

\end{figure*}
Figure~\ref{fig:hilo-muaompc} depicts an overview of the embedded module. 
Similarly to other  modules, the \mintinline{python}{LMPC} class is initialized 
using a \mintinline{python}{Model} object. In this case, a linear time-invariant discrete-time
model is used. Calling the \mintinline{python}{LMPC.setup()} method, specifying the solver 
\mintinline{python}{'muaompc'}, will trigger the code generation procedure.
The code generation automatically creates C-code which can be compiled
to run on an embedded target. Additionally, a Python interface to the
generated C-code is also created. This Python interface is used by the \mintinline{python}{LMPC.optimize()} method to obtain the input that minimizes \ref{eqn:mpcproblem}. The basic use of the Embedded Module is exemplified in the following Python code

\begin{minted}[bgcolor=bg]{python}

from hilo_mpc import LMPC, Model

# The Model class uses the linear system matrices A, B, and 
# the LMPC class uses the weighting matrices Q, R, P, constraints, etc.,
# to create the mpc object. Using the solver muaompc triggers the code generation
mpc.setup(nlp_solver='muaompc')
# the generated MPC code can be used directly on an embedded target using C,
# or via the Python interface as HILO-MPC does on the next line
u = mpc.optimize(x0=x)
\end{minted}

\section{Examples}
\label{sec:examples}

We underline the flexibility of the toolbox considering a series of examples, which can be also found in the repository, as Python files or Jupyter notebooks. Thus, we do not provide implementation details, just a high-level description of the problems and outline some of the results.

\subsection{Learning the System Dynamics -  Racing a Minicar}
\label{sec:example-race-car}
\begin{figure*}[htb]%{.49\linewidth}
    \captionsetup[sub]{justification=centering,font=large}
    \centering
    \begin{subfigure}[t]{.49\linewidth}
        \inputpdf{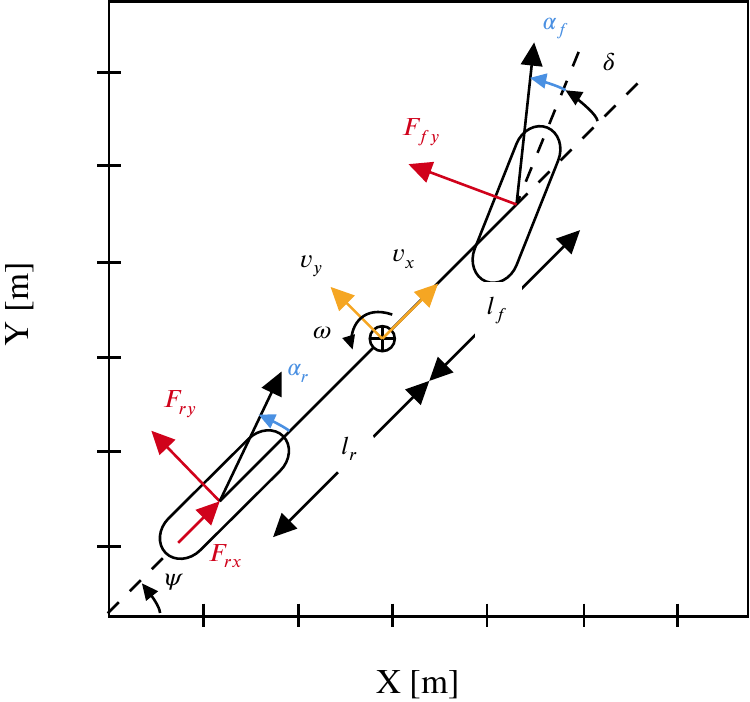}
        \caption{Complex single-track/bicycle car model.}
        \label{fig:bike1}
    \end{subfigure}
    \begin{subfigure}[t]{.49\linewidth}
        \inputpdf{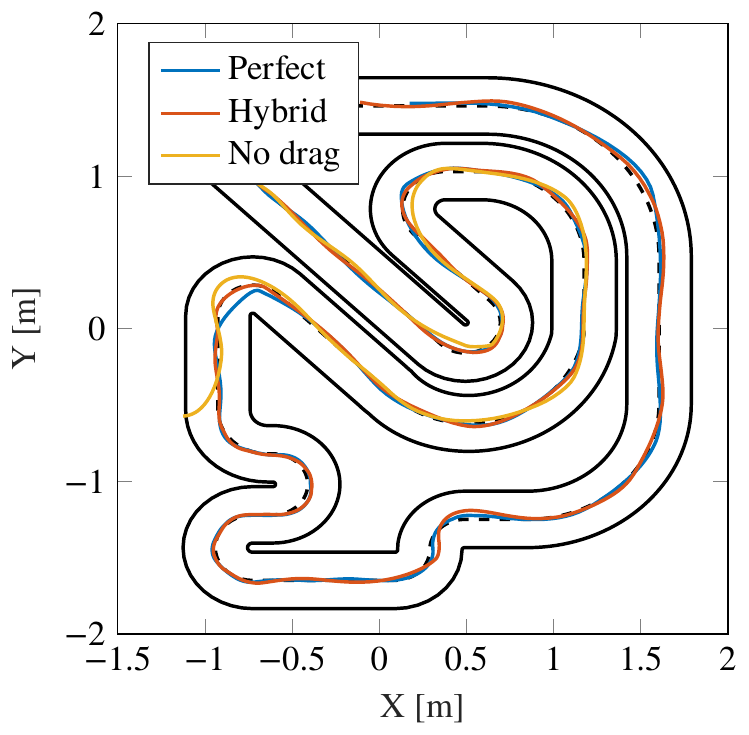}
        \caption{Race track and results.}
        \label{fig:race_car}
    \end{subfigure}
    \caption{Example of a racing car driving on a racetrack following a given path.}
\end{figure*}

We consider autonomous racing of a minicar, based on the results and models presented in Liniger~et~al.\cite{Liniger2015} The goal is to use model predictive control to follow a complex track using a path following formulation in presence of disturbances (see Figure \ref{fig:bike1}). An extended single-track model of the car, see \ref{fig:bike1} and section \ref{sec:modeling_module}, including nonlinear tire models and drive train models as identified and validated in experiments in Liniger~et~al.\cite{Liniger2015} is used. We extend the model adding a component describing the effect of lateral drag forces due to, for example, a strong wind. The overall model is represented by the following system of nonlinear differential equations
\begin{align*}
\dot{p}_x &= v_x \cos (\psi) - v_y \sin(\psi), \quad%\
\dot{p}_y = v_x \sin (\psi) + v_y \cos(\psi), \quad
\dot{\psi} = w, \\
\dot{v}_x &= \frac{1}{m}\left( F_{r,x}-F_{a,x}- (F_{f,y}-F_{a,y}) \sin(\delta) + m v_y \omega \right), \quad
\dot{v}_y = \frac{1}{m}\left( F_{r,y}-F_{a,y}- (F_{f,y}-F_{a,y}) \cos(\delta) - m v_x \omega \right), \\
\dot{\omega} &= \frac{1}{I_z} \left( F_{f,y}l_f \cos(\delta) - F_{r,y}l_r\right),
\end{align*}
where $p_x$ and $p_y$ are the coordinates of the center of gravity, $v_x$ and $v_y$ are longitudinal and lateral velocities of the center of gravity. The orientation is denoted by $\psi$ and the yaw rate by $\omega$. The control inputs are the motor duty cycle $d$ and steering angle $\delta$. The two parameters $l_f$ and $l_r$ are the distances from the center of gravity to the front axle and rear axle, respectively, $m$ is the mass of the car, and $I_z$ is the inertia.  Figure~\ref{fig:bike1} shows the single-track car model. The path is the center of a racing track that has been interpolated using splines.
The tire forces are modeled with a simpliﬁed Pacejka Tire Model\cite{bakker1987tyre}
%\begin{subequations*}
\begin{align*}
F_{f,y} &= D_f \sin(C_f \arctan(B_f \alpha_f)),\quad
F_{r,y} = D_r \sin(C_r \arctan(B_r \alpha_r)), \quad
F_{r,x} = (C_{m1}-C_{m2}v_x)d - C_r - C_d v^2_{x}, \\
\alpha_f &= - \arctan\left( \frac{w l_f + v_y}{v_x} \right) + \delta,\quad
\alpha_r = \arctan\left( \frac{w l_r - v_y}{v_x} \right)
\end{align*}
%\end{subequations*}
where $D_f$, $D_r$, $B_f$, $B_r$, $C_{m1}$, $C_{m2}$, $C_r$, and $C_d$ are parameters. The longitudinal and lateral drag forces are defined, respectively, as 
\begin{align*}
    F_{a,x} = 0.5 \,\  c_w  \rho  A  v_x, \quad
    F_{a,y} = 0.5 \,\ c_w  \rho  A  v_{y,\text{wind}}, 
\end{align*}
where $c_w$ is the drag coefficient, $\rho$ is the air density, $A$ is the effective flow surface, and $v_{y,wind}$ is the lateral wind velocity.
The model used by the MPC does not have the drag effect. The goal is to learn this effect from data using a neural network and then augment the first principles model with a machine learning component that models the drag effect. After discretization, the hybrid model can be written as
\begin{equation}
    x_{k+1} = f(x_k,u_k) + B^\Tr m(x_k,u_k),
\end{equation}
where $m(x_k,u_k)$ is an NN model. The features of the NN are $v_{y,\text{wind}}$ and $v_x$, the labels are correction terms for $\phi$, $v_x$, $v_y$, and $\omega$ via the weighting matrix $B^\Tr$. To show the effectiveness of the learning, we compare the results of MPC using the perfect model (i.e., with known drag force effects), the hybrid model (using the NN), and the model without drag forces. Furthermore, the measurements of position, velocity, and directions are affected by Gaussian noise and estimated using an unscented KF.
Figure \ref{fig:race_car} shows the results of the simulation. While the hybrid model has similar results as the perfectly-known model, the model without drag exits the race area after the fifth curve. %The complete code can be found in the HILO-MPC repository.

\subsection{Learning a Reference - Cooperative Robots}
\label{sec:cooperative-robots}
\begin{figure*}[htb]
    \centering
    \inputpdf{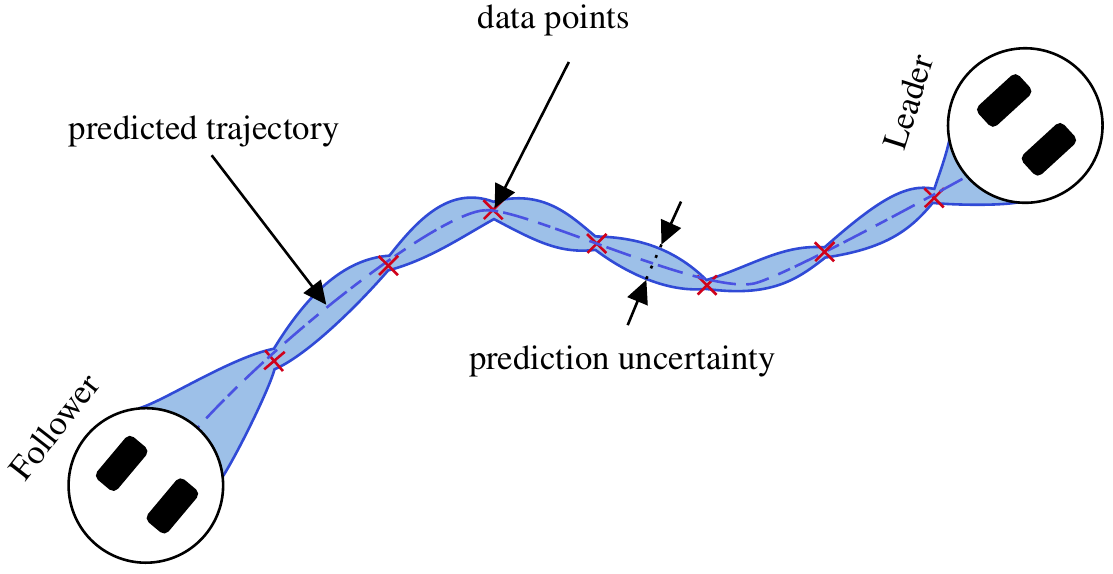}
    \caption{Cooperative robot example. The follower should track the leader, who follows a periodic, yet unknown trajectory. The trajectory is learned from data using a GP.}
    \label{fig:mobile-robots}
\end{figure*}

Inspired by\cite{Matschek2021a} we consider that a follower robot has to track the position of a leader robot, where the leader moves along a periodic but unknown trajectory (see Figure~\ref{fig:mobile-robots}). 
 
\begin{figure*}[htb]
    \centering
    \inputpdf{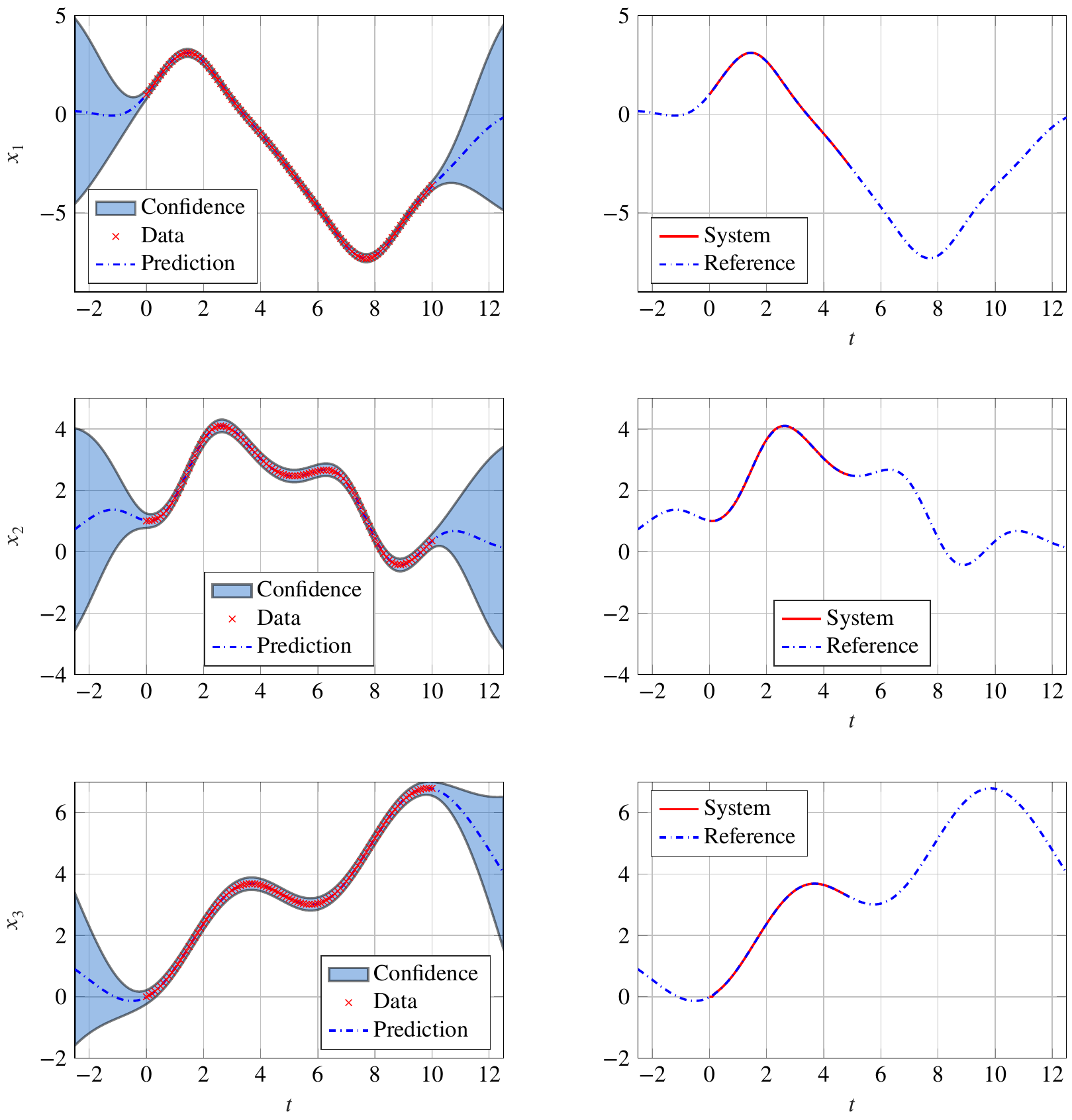}
    \caption{Left column: learned GP. Right column: states of the follower robot following the learned references.}
    \label{fig:mobile-robots-results}
\end{figure*}

 The objective is to learn the trajectory of the leader with GPs, which are used in the follower. Hence, in this case the machine learning model enters the cost function. The nonlinear dynamics of the robots are described with the following simple dynamics ODE
%\begin{subequations}
    \begin{align*}
    \dot{x}_1 &= u_1 \sin(x_3), \\
    \dot{x}_2 &= u_1 \cos(x_3), \\
    \dot{x}_3 &= u_2.
    \end{align*}
%\end{subequations}
Here $x_1$ and $x_2$ are the horizontal and vertical position of the root, $x_3$ its heading angle, $u_1$ is the speed, and $u_2$ the turning rate.
The resulting tracking problem takes the form \eqref{eq:problem-mpc} with the objective function \eqref{eq:mpc-traj-tracking}, where $x_r(\cdot)$ is the mean function of a GP trained on the data set collected from the position of the leader.
The trajectory generated by the leader results from applying the time-varying forces
%\begin{subequations}
%    \begin{align}
 $u_1(t) = 2 + 0.1 \sin(t)$ and $u_2(t) = 0.5 + \sin(t)$.
%    \end{align}
%\end{subequations}
Figure \ref{fig:mobile-robots-results} shows the results of the reference learning and closed-loop simulations.

\subsection{Learning the Controller - Spring Damper System}
\label{sec:spring-damper}
\begin{figure*}[htb]
    \centering
    \inputpdf{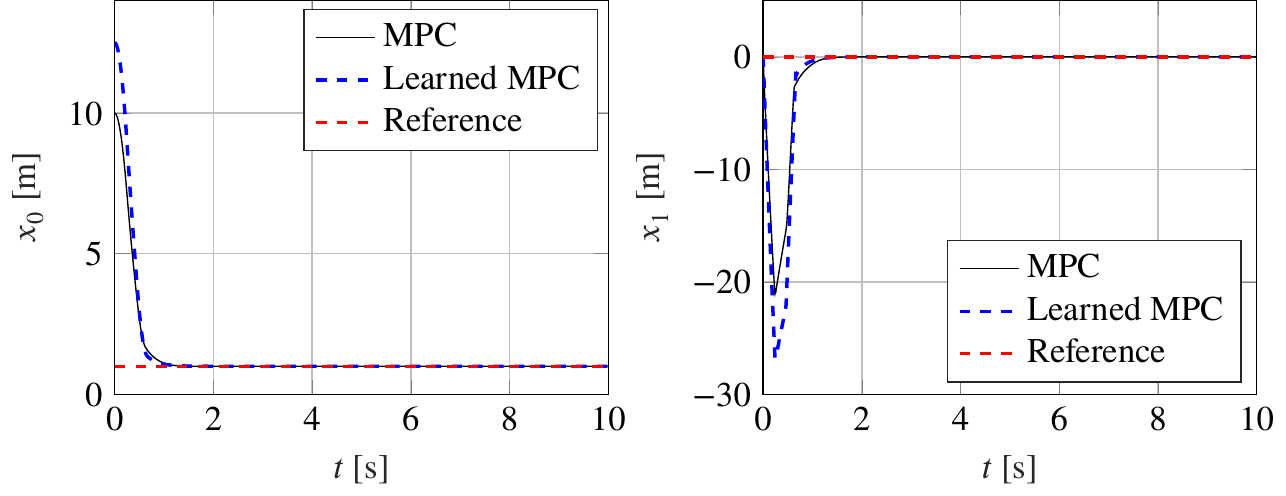}
    \caption{Comparison of the learned and exact MPC.}
    \label{fig:ml-controller-results}
\end{figure*}

Solving an MPC requires the solution of a (nonlinear) optimization problem online. 
%This is possible only for applications where the available computational power and energy are sufficient to guarantee a solution within the sampling times. 
For embedded applications with low computational power or applications that can use only a limited amount of energy (for example battery powered systems) this is often not possible. Hence, methods that provide at least a close-to-optimal solution without solving the optimization problem at every time step are necessary. This can be done using \emph{explicit} MPC approaches.\cite{alessio2009survey,besselmann2012explicit} One way to avoid the explicit solution is to learn the solution of an MPC offline, i.e., to learn the map $x \mapsto \rho_{\theta}(x)$ that (approximates) the implicit MPC control law, and then using the learned controller online (see Figure~\ref{fig:scheme}).\cite{Parisini1995,Karg2020,Maddalena2020,Cao2020,Chen2018,Cseko2015,Ponkumar2018}
In this way, the control action can be found with a simple and fast function evaluation.
We consider the control of a mass-spring-damper system using a learned controller. The models is given by
\begin{subequations}
    \begin{align}
& \dot{x}_1 = x_2, \\
& \dot{x}_2 = \frac{1}{m} (u - k x_1 - d x_2) ,
    \end{align}
\end{subequations}
where $x_1$ is the vertical position, $x_2$ the vertical velocity, $u$ the vertical force, and $k$, $d$ the system parameters. The equilibrium point is $x=(0,0)$, $u=0$. The objective is to maintain the reference $x_{\text{ref}}=[1,0]$ using a learned MPC. 
%To do so we use the results of just one closed-loop MPC simulation starting from the initial conditions $x(0)=(12,0)$. 
In total 667 data points, obtained by simulations, are collected. We use the data to train an NN via HILO-MPC with three fully-connected layers, with 10 neurons each. The features of the NN are $x_1$ and $x_2$, and the label is the input $u$. We test the learned controller starting from a different initial condition $x(0)=(10,0)$. In Figure~\ref{fig:ml-controller-results} the simulation results are shown. The learned controller is able to bring the system to the reference like the original controller.

\subsection{Embedded MPC - HIL Simulation using an Arduino}
\label{sec:embedded}
We conclude by presenting a hardware-in-the-loop (HIL) simulation 
based on the code generated by the embedded module (see 
Subsection~\ref{sec:emb-mod}).

Figure~\ref{fig:hilo-muaompc} depicts the code generation and HIL simulation setup.
The generated C-code is cross compiled to run on an Arduino 
microcontroller. 
The system dynamics are simulated on a laptop computer using the nominal linear model $x^+=Ax+Bu$.
%For any given input $u$, and using the current state $x$,
%the simulation computes the state at the next sampling period $x^+$.
Using a serial interface (i.e. USB to UART), the state $x^+$
is transferred to the Arduino, 
%which uses it as parameter $x$
%to solve the MPC problem, i.e., 
to compute a new sequence $\vb{u}$
by solving the parametric QP $\mathcal{P}(x)$.
Only the first part $u_0$ of this sequence is transferred by the Arduino to the computer via the serial interface.%, and thus closing the loop.

The system to be controlled is the 
linearized %version of the race car. 
%We assume the car is traveling at constant speed, 
%and only consider the 
lateral dynamics of the race car.
The input is the steering angle $u=\delta$.
The state is defined as $x = [p_y \ \ \psi \ \ \omega]^Tr$,
where $p_y$ is the lateral position of the center of mass of the car,
and $\psi$ is the orientation
of the car with respect to the $X$-axis and $\omega = \dot \psi$
(refer to Figure~\ref{fig:bike1}).
The input is constrained to $-0.1 \leq u \leq 0.1$,
and the state $x_1 = \psi$ must satisfy $-0.1 \leq x_1 \leq 0.1$.
%The objective is to bring the state to the origin, i.e. $x = \vb{0}$,
%while satisfying input and state constraints.

For the HIL simulation, we used an Arduino Nano 33 BLE.\cite{arduinonano} 
\footnote{It has a 32-bit Cortex-M4F microcontroller core (with single precision floating-point unit) 
running at 64 MHz.} 
We use a discretization time of $15\,\text{ms}$, and a horizon length of $20$ steps.
The Arduino requires around $10\,\text{ms}$ to find (an approximate) solution 
to the MPC problem on each iteration. The compiled C-code takes about
$20\,\text{kB}$ of flash memory.

\begin{figure*}[htb]
  \centering
  \inputpdf{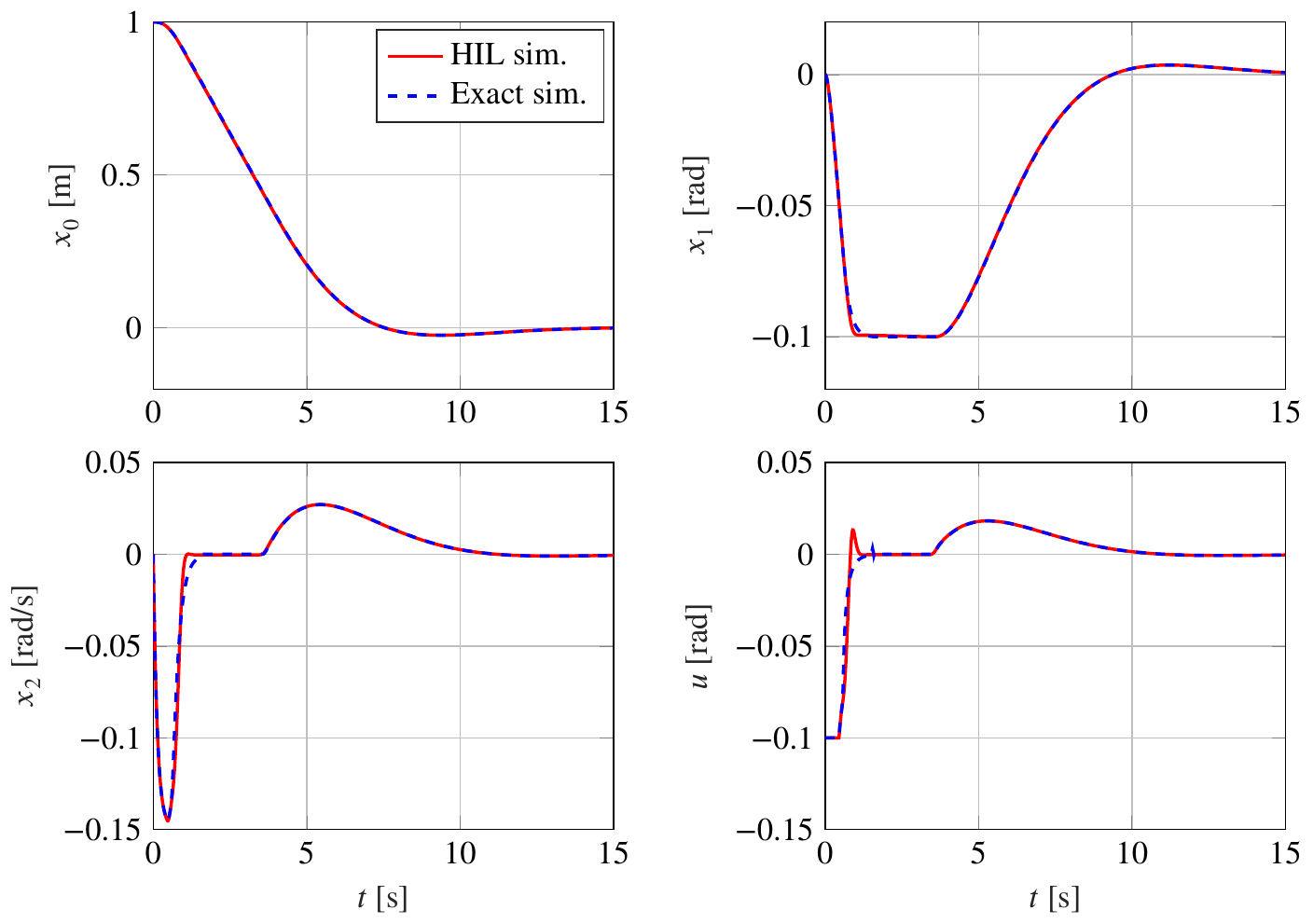}
  \caption{Comparison of HIL simulation and exact simulation}
  \label{fig:hil-sim-plot}
\end{figure*}

Figure~\ref{fig:hil-sim-plot} shows the compared trajectories of states and input
for the HIL simulation (approximate solution on Arduino using single precision float)
and the exact solution computed by a laptop using HILO-MPC's
interface to CPLEX. 
%In the figures, the difference between exact and approximate solution
%are not significant, except where the state $x_1$ is close
%to its constraint (around $0.1$ seconds).
Note that neither the input nor the state constraints are violated. 
The main differences are due to the early
termination of the embedded algorithm, and the use of 
single precision float.

\section{Conclusions and Outlook}
\label{sec:conclusions}
While the use of machine learning approaches for model-based optimal control, such as predictive control and estimation are becoming increasingly important, currently there is no easy-to-use open-source tool available to support the researcher and developer. We outlined the concept and ideas behind HILO-MPC, a toolbox for fast development of optimal and predictive control and estimation methods that facilitates the use of machine learning models trained using PyTorch and TensorFlow for simulations and embedded applications. The concept is easily expandable and we underlined the simple use considering four examples.
The toolbox allows to consider different methods, spanning from model predictive and optimal control, moving horizon estimation, and Kalman filters to particle filters. We believe that it is a flexible, yet easy-to-use tool for research and teaching. 
% We are currently working on a fifth module that allows solving optimization problems efficiently and fast, aiming at embedded applications such as embedded MPC. 
Future developments will allow using extended machine learning models such as recurrent neural networks, Bayesian neural networks and reinforcement learning, as well as extended control formulations, such as tube-based MPC, stochastic MPC, \cite{Mayne2005} and multi-mode MPC.\cite{bethge2018multi, morabito2019multi} Furthermore, we are expanding the toolbox towards code generation for embedded nonlinear MPC using the Al'brekht's method.\cite{Kallies2020a,Kallies2020b,Kallies2021} 

\section*{Acknowledgements}  % \acks command that should be used according to the manual is not working

The authors thank Johanna Bethge, Michael Maiworm, Tim Zieger, Maik Pfefferkorn, Rudolph Kok, Sebasti{\'a}n Espinel-R{\'i}os and Janine Matschek for providing feedback and examples for HILO-MPC. Special thanks go to Lena Kranert for the cricket.

\bibliography{references}

\end{document}

%% file: colors.tex
\definecolor{muaompccolor}{RGB}{254,246,233}
\definecolor{embeddedcolor}{RGB}{248,193,103}
\definecolor{controlcolor}{RGB}{217,129,92}
\definecolor{modelcolor}{RGB}{232,103,79}
\definecolor{machinelearningcolor}{RGB}{208, 2, 27}
\definecolor{estimationcolor}{RGB}{207,123,123}
\definecolor{uncertaintycolor}{RGB}{11,97,198}
\definecolor{bg}{rgb}{0.95,0.95,0.95}

%% file: figures/control_loop.tex
\tikzset{every picture/.style={line width=0.75pt}} %set default line width to 0.75pt        

\begin{tikzpicture}[x=0.75pt,y=0.75pt,yscale=-1,xscale=1]

%Shape: Rectangle [id:dp40927779822341837] 
\draw  [fill={rgb, 255:red, 208; green, 2; blue, 27 }  ,fill opacity=0.7 ] (439.83,52.53) -- (552.66,52.53) -- (552.66,86.4) -- (439.83,86.4) -- cycle ;
%Shape: Rectangle [id:dp9580357372458703] 
\draw   (439.83,125.35) -- (552.66,125.35) -- (552.66,159.22) -- (439.83,159.22) -- cycle ;
%Shape: Rectangle [id:dp1159723749066921] 
\draw  [fill={rgb, 255:red, 207; green, 123; blue, 123 }  ,fill opacity=0.7 ] (439.83,196.48) -- (552.66,196.48) -- (552.66,230.35) -- (439.83,230.35) -- cycle ;
%Shape: Rectangle [id:dp9349588461557252] 
\draw  [fill={rgb, 255:red, 217; green, 129; blue, 92 }  ,fill opacity=0.7 ] (80.33,88.54) -- (293.97,88.54) -- (293.97,195.77) -- (80.33,195.77) -- cycle ;
%Shape: Rectangle [id:dp21201448083288543] 
\draw  [fill={rgb, 255:red, 255; green, 255; blue, 255 }  ,fill opacity=1 ] (88.74,114) -- (185,114) -- (185,147.87) -- (88.74,147.87) -- cycle ;

%Shape: Rectangle [id:dp5178566860414551] 
\draw  [fill={rgb, 255:red, 255; green, 255; blue, 255 }  ,fill opacity=1 ] (193.08,114) -- (289.33,114) -- (289.33,147.87) -- (193.08,147.87) -- cycle ;

%Shape: Rectangle [id:dp47627104059615477] 
\draw  [fill={rgb, 255:red, 255; green, 255; blue, 255 }  ,fill opacity=1 ] (89.08,156.34) -- (185.33,156.34) -- (185.33,190.21) -- (89.08,190.21) -- cycle ;

%Shape: Rectangle [id:dp5152894328953086] 
\draw  [fill={rgb, 255:red, 255; green, 255; blue, 255 }  ,fill opacity=1 ] (193.08,156.34) -- (289.33,156.34) -- (289.33,190.21) -- (193.08,190.21) -- cycle ;

%Shape: Rectangle [id:dp018822904302240318] 
\draw  [fill={rgb, 255:red, 248; green, 193; blue, 103 }  ,fill opacity=0.7 ] (134.05,212.15) -- (246.87,212.15) -- (246.87,246.02) -- (134.05,246.02) -- cycle ;

%Shape: Rectangle [id:dp46125286852646563] 
\draw  [fill={rgb, 255:red, 208; green, 2; blue, 27 }  ,fill opacity=0.7 ] (133.38,44.19) -- (246.2,44.19) -- (246.2,78.06) -- (133.38,78.06) -- cycle ;

%Straight Lines [id:da5241374195313209] 
\draw    (296.79,141.09) -- (434.99,141.09) ;
\draw [shift={(437.99,141.09)}, rotate = 180] [fill={rgb, 255:red, 0; green, 0; blue, 0 }  ][line width=0.08]  [draw opacity=0] (6.25,-3) -- (0,0) -- (6.25,3) -- cycle    ;
%Straight Lines [id:da006180510898959435] 
\draw  [dash pattern={on 0.84pt off 2.51pt}]  (375.67,60.25) -- (375.67,138.58) ;
\draw [shift={(375.67,141.58)}, rotate = 270] [fill={rgb, 255:red, 0; green, 0; blue, 0 }  ][line width=0.08]  [draw opacity=0] (6.25,-3) -- (0,0) -- (6.25,3) -- cycle    ;
%Straight Lines [id:da6415183479928961] 
\draw  [dash pattern={on 0.84pt off 2.51pt}]  (375.67,229.74) -- (375.67,144.58) ;
\draw [shift={(375.67,141.58)}, rotate = 90] [fill={rgb, 255:red, 0; green, 0; blue, 0 }  ][line width=0.08]  [draw opacity=0] (6.25,-3) -- (0,0) -- (6.25,3) -- cycle    ;
%Straight Lines [id:da6951394694577886] 
\draw  [dash pattern={on 0.84pt off 2.51pt}]  (246.26,59.98) -- (374.66,59.98) ;
%Straight Lines [id:da8431518411243517] 
\draw  [dash pattern={on 0.84pt off 2.51pt}]  (106.59,28.11) -- (106.59,84.44) ;
\draw [shift={(106.59,87.44)}, rotate = 270] [fill={rgb, 255:red, 0; green, 0; blue, 0 }  ][line width=0.08]  [draw opacity=0] (6.25,-3) -- (0,0) -- (6.25,3) -- cycle    ;
%Straight Lines [id:da1485037143270571] 
\draw  [dash pattern={on 0.84pt off 2.51pt}]  (106.26,27.26) -- (496.66,27.26) ;
%Straight Lines [id:da05244399077706552] 
\draw  [dash pattern={on 0.84pt off 2.51pt}]  (496.25,52.39) -- (496.25,27.26) ;
%Straight Lines [id:da1784106810550885] 
\draw    (402.08,77.51) -- (435.34,77.51) ;
\draw [shift={(438.34,77.51)}, rotate = 180] [fill={rgb, 255:red, 0; green, 0; blue, 0 }  ][line width=0.08]  [draw opacity=0] (6.25,-3) -- (0,0) -- (6.25,3) -- cycle    ;
%Straight Lines [id:da9843458897800805] 
\draw    (402.08,77.51) -- (402.08,141.09) ;

%Straight Lines [id:da5429100261096111] 
\draw    (590.01,205.17) -- (556.74,205.16) ;
\draw [shift={(553.74,205.16)}, rotate = 0.02] [fill={rgb, 255:red, 0; green, 0; blue, 0 }  ][line width=0.08]  [draw opacity=0] (6.25,-3) -- (0,0) -- (6.25,3) -- cycle    ;
%Straight Lines [id:da9490614061978668] 
\draw    (590.01,204.85) -- (590.01,141.94) ;
%Straight Lines [id:da24172364273970337] 
\draw    (554.19,141.94) -- (590.01,141.94) ;
%Straight Lines [id:da5496609050981129] 
\draw    (590.01,141.94) -- (590.01,77.86) ;
%Straight Lines [id:da5103992322352211] 
\draw    (590.42,77.86) -- (570.48,77.85) -- (557.16,77.85) ;
\draw [shift={(554.16,77.84)}, rotate = 0.02] [fill={rgb, 255:red, 0; green, 0; blue, 0 }  ][line width=0.08]  [draw opacity=0] (6.25,-3) -- (0,0) -- (6.25,3) -- cycle    ;
%Straight Lines [id:da9625248045417865] 
\draw    (50.66,261.15) -- (496.9,261.15) ;
%Straight Lines [id:da5169697615021076] 
\draw    (496.25,229.36) -- (496.25,261.59) ;
%Straight Lines [id:da770181023387118] 
\draw  [dash pattern={on 0.84pt off 2.51pt}]  (552.66,61.14) -- (602.66,61.14) ;
%Straight Lines [id:da535359715167171] 
\draw  [dash pattern={on 0.84pt off 2.51pt}]  (602.66,221.22) -- (602.66,61.14) ;
%Straight Lines [id:da7017013305970756] 
\draw  [dash pattern={on 0.84pt off 2.51pt}]  (555.66,221.22) -- (602.66,221.22) ;
\draw [shift={(552.66,221.22)}, rotate = 0] [fill={rgb, 255:red, 0; green, 0; blue, 0 }  ][line width=0.08]  [draw opacity=0] (6.25,-3) -- (0,0) -- (6.25,3) -- cycle    ;
%Straight Lines [id:da8027503847635218] 
\draw    (402.67,203.84) -- (435.74,203.84) ;
\draw [shift={(438.74,203.84)}, rotate = 180] [fill={rgb, 255:red, 0; green, 0; blue, 0 }  ][line width=0.08]  [draw opacity=0] (6.25,-3) -- (0,0) -- (6.25,3) -- cycle    ;
%Straight Lines [id:da6346714991836258] 
\draw    (402.08,204) -- (402.08,141.09) ;
%Straight Lines [id:da585265857065806] 
\draw  [dash pattern={on 0.84pt off 2.51pt}]  (246.56,230.3) -- (375.33,230.3) ;
%Straight Lines [id:da704282652926296] 
\draw  [dash pattern={on 0.84pt off 2.51pt}]  (189.33,195.74) -- (189.33,208.4) ;
\draw [shift={(189.33,211.4)}, rotate = 270] [fill={rgb, 255:red, 0; green, 0; blue, 0 }  ][line width=0.08]  [draw opacity=0] (6.25,-3) -- (0,0) -- (6.25,3) -- cycle    ;
%Straight Lines [id:da9286800597156812] 
\draw  [dash pattern={on 0.84pt off 2.51pt}]  (190.59,27.44) -- (190.59,40.11) ;
\draw [shift={(190.59,43.11)}, rotate = 270] [fill={rgb, 255:red, 0; green, 0; blue, 0 }  ][line width=0.08]  [draw opacity=0] (6.25,-3) -- (0,0) -- (6.25,3) -- cycle    ;
%Straight Lines [id:da589121137669645] 
\draw    (50.66,261.15) -- (50.66,59.59) ;
%Straight Lines [id:da2700689464237993] 
\draw    (51.99,150.37) -- (76.02,150.37) ;
\draw [shift={(79.02,150.37)}, rotate = 180] [fill={rgb, 255:red, 0; green, 0; blue, 0 }  ][line width=0.08]  [draw opacity=0] (6.25,-3) -- (0,0) -- (6.25,3) -- cycle    ;
%Straight Lines [id:da9771297559989931] 
\draw    (51.99,229.71) -- (130.99,229.71) ;
\draw [shift={(133.99,229.71)}, rotate = 180] [fill={rgb, 255:red, 0; green, 0; blue, 0 }  ][line width=0.08]  [draw opacity=0] (6.25,-3) -- (0,0) -- (6.25,3) -- cycle    ;
%Straight Lines [id:da1552607281243097] 
\draw    (50.66,59.59) -- (129.66,59.59) ;
\draw [shift={(132.66,59.59)}, rotate = 180] [fill={rgb, 255:red, 0; green, 0; blue, 0 }  ][line width=0.08]  [draw opacity=0] (6.25,-3) -- (0,0) -- (6.25,3) -- cycle    ;

% Text Node
\draw (496.25,69.46) node  [font=\footnotesize] [align=left] {Machine \\learning};
% Text Node
\draw (496.25,142.28) node  [font=\footnotesize] [align=left] {Plant};
% Text Node
\draw (496.25,213.41) node  [font=\footnotesize] [align=left] {Estimator};
% Text Node
\draw (358.72,11.99) node [anchor=north west][inner sep=0.75pt]   [align=left] {{\footnotesize model}};
% Text Node
\draw (561.42,44.73) node [anchor=north west][inner sep=0.75pt]   [align=left] {{\footnotesize model}};
% Text Node
\draw (284.98,245.73) node [anchor=north west][inner sep=0.75pt]  [font=\footnotesize] [align=left] {estimates};
% Text Node
\draw (321.64,125.8) node [anchor=north west][inner sep=0.75pt]  [font=\footnotesize] [align=left] {input};
% Text Node
\draw (506.44,99) node [anchor=north west][inner sep=0.75pt]  [font=\footnotesize] [align=left] {measurements};
% Text Node
\draw (137.2,130.93) node  [font=\footnotesize] [align=left] {Objective function};
% Text Node
\draw (241.2,173.28) node  [font=\footnotesize] [align=left] {References};
% Text Node
\draw (241.2,130.93) node  [font=\footnotesize] [align=left] {Constraints};
% Text Node
\draw (137.2,173.28) node  [font=\footnotesize] [align=left] {Dynamic model};
% Text Node
\draw (131.72,96.33) node [anchor=north west][inner sep=0.75pt]  [font=\footnotesize] [align=left] {MPC / Optimal control};
% Text Node
\draw (190.46,229.08) node  [font=\footnotesize] [align=left] {Embedded MPC};
% Text Node
\draw (189.79,61.13) node  [font=\footnotesize] [align=left] {Machine learning \\controller};

\end{tikzpicture}